\documentclass[pdflatex,sn-nature]{sn-jnl}

\usepackage{graphicx}%
\usepackage{multirow}%
\usepackage{amsmath,amssymb,amsfonts}%
\usepackage{amsthm}%
\usepackage{mathrsfs}%
\usepackage[title]{appendix}%
\usepackage{xcolor}%
\usepackage{textcomp}%
\usepackage{manyfoot}%
\usepackage{booktabs}%
\usepackage{algorithm}%
\usepackage{algorithmicx}%
\usepackage{algpseudocode}%
\usepackage{listings}%

\theoremstyle{thmstyleone}%
\theoremstyle{thmstyletwo}%

\theoremstyle{thmstylethree}%

\begin{document}

\title[Article Title]{Longwang: Zero-Shot Global Spatiotemporal Precipitation Downscaling with a Latent Generative Prior}


\author*[1]{\fnm{Yue} \sur{Wang}}\email{yw2796@cornell.edu}

\author[1]{\fnm{Daniele} \sur{Visioni}}\email{dv224@cornell.edu}

\affil*[1]{\orgdiv{Department of Earth and Atmospheric Sciences}, \orgname{Cornell University}, \orgaddress{\street{112 Hollister Drive}, \city{Ithaca}, \postcode{14853}, \state{NY}, \country{USA}}}

\abstract{High-resolution precipitation information is essential for climate impact assessment, yet global climate models remain too coarse to resolve key small-scale processes. Existing machine learning downscaling methods often require paired low- and high-resolution data for supervised learning, are tied to fixed regions or scale factors during inference, and can be computationally expensive to train and run in physical space. Here we introduce Longwang, a zero-shot latent generative framework for global spatiotemporal precipitation downscaling. Longwang learns a context-conditioned latent generative prior and combines it with a physically informed observation operator through posterior sampling, enabling daily $\mathcal{O}(10\,\mathrm{km})$ precipitation fields to be generated from monthly $\mathcal{O}(100\,\mathrm{km})$  inputs. On ERA5 reanalysis, Longwang outperforms standard posterior sampling with an unconditional generative prior in reconstructing fine-scale spatial patterns, preserving temporal coherence, and recovering extreme precipitation intensities. The framework further generalizes to historical climate simulations and future climate projections under substantial distribution shift.}

\maketitle

\section{Introduction}
\label{sec:intro}

Anthropogenic climate change is increasing the frequency and severity of many extreme weather events \cite{zhang2024anthropogenic}, underscoring the need for high-resolution climate data to assess future impacts. However, global climate simulations used for such projections are typically run at coarse ($\sim$ 100km) spatial resolutions because of computational constraints~\cite{schneider2017climate}. Statistical downscaling has therefore become an important approach for bridging this resolution gap without relying on similarly expensive dynamical downscaling methods \cite{giorgi2015regional}. Precipitation, which directly affects water resources, agriculture, and flood risk~\cite{kotz2022effect}, is particularly challenging to downscale because of its spatial intermittency, temporal variability, and heavy-tailed intensity distribution. Capturing this regional and temporal heterogeneity within a unified global downscaling framework at high spatiotemporal resolution remains an open challenge.

\begin{figure}[tbp]
\centering
\includegraphics[width=\textwidth]{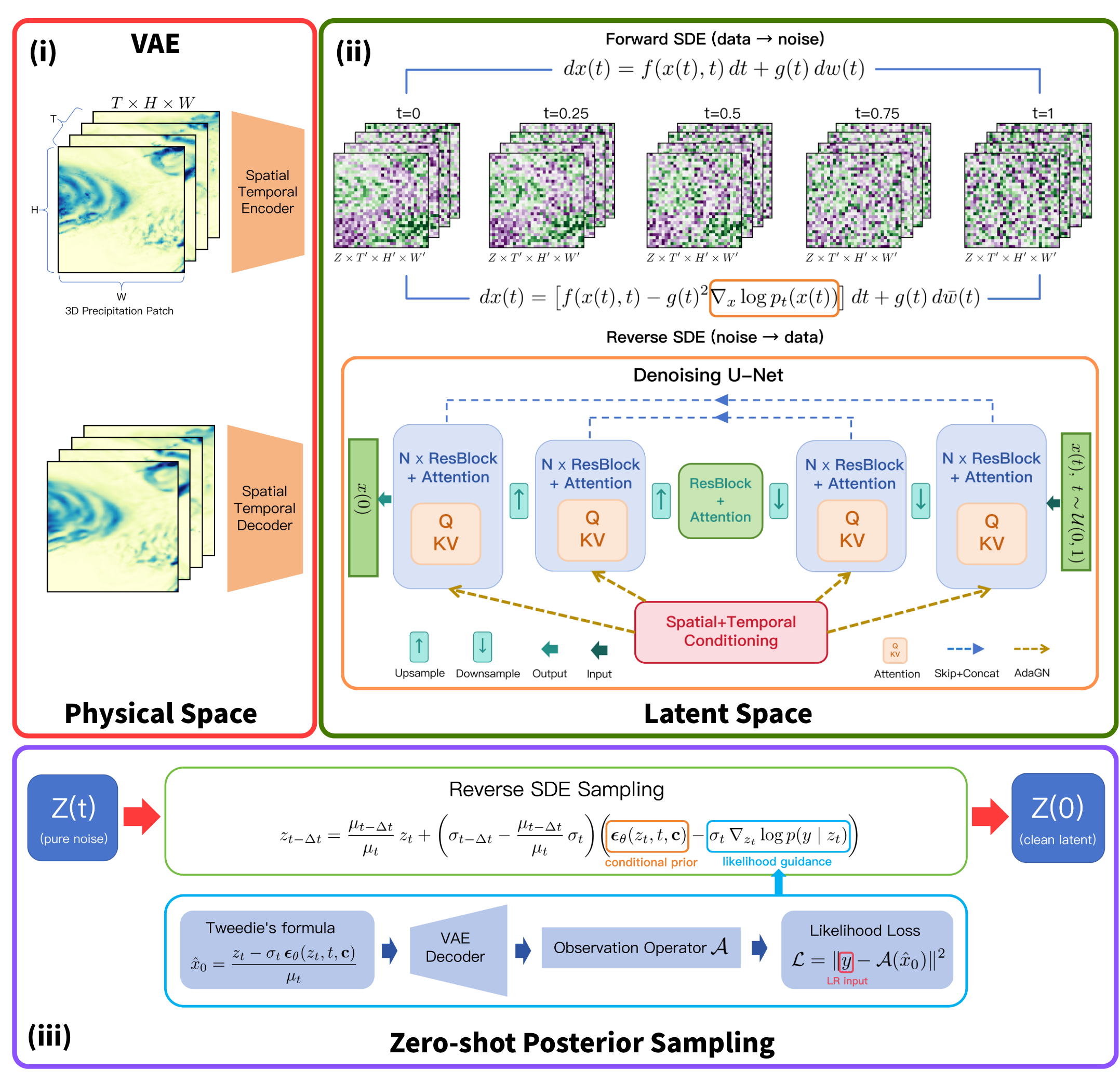}
\caption{\textbf{Schematic overview of Longwang.} \textbf{(i)} Longwang encodes high-resolution spatiotemporal precipitation fields into a compact latent space using a variational autoencoder (VAE), with a decoder reconstructing fields in physical space. \textbf{(ii)} A score-based diffusion model is trained in this latent space, where the forward stochastic differential equation (SDE) progressively adds Gaussian noise and a denoising U-Net learns the reverse dynamics. Spatiotemporal context conditions the latent generative prior. \textbf{(iii)} During zero-shot posterior downscaling, reverse-SDE sampling is guided by the likelihood of the observed low-resolution precipitation field. At each step, Tweedie’s formula estimates the clean latent state, which is decoded and passed through the observation operator to enforce consistency with the low-resolution input.}
\label{fig:schematic}
\end{figure}

Statistical downscaling based on deep learning has become a common approach for this task. Convolutional neural networks trained with deterministic regression objectives can produce high-resolution fields with realistic fine-scale structures~\cite{hess2022deep,bano2020configuration,wang2026mitigating,wang2021deep}. However, such models typically return a single estimate and therefore cannot represent uncertainty in the non-unique mapping from coarse- to fine-resolution climate fields. Generative methods address this limitation by learning distributions over high-resolution fields, yielding multiple plausible realizations and enabling uncertainty quantification through sampling~\cite{harris2022generative,mardani2025residual,hess2025fast}.

Most generative downscaling studies to date have focused on spatial downscaling alone, including methods trained on real climate data~\cite{harris2022generative,mardani2025residual,hess2025fast} and idealized fluid systems~\cite{wan2023debias,bischoff2024unpaired}. However, applying spatial downscaling independently at each time step does not explicitly model temporal dependence, which is important for representing coherent high-resolution sequences. More recently, several studies have extended generative downscaling to joint spatiotemporal settings~\cite{srivastava2024precipitation,glawion2025global,schmidt2025generative}. Many of these approaches learn supervised conditional mappings from paired coarse- and fine-resolution data~\cite{srivastava2024precipitation,glawion2025global}. This design ties the trained model to the coarse input resolution and coarsening operator used during training, and typically requires retraining when the downscaling setting changes.

An alternative is to learn a generative prior over high-resolution fields and, at inference time, combine it with a likelihood score through Bayesian posterior sampling~\cite{rozet2023score}. By decoupling the learned prior from the observation model, this formulation enables zero-shot application to new downscaling configurations without retraining the prior. Existing approaches, however, define the prior directly in the full high-resolution physical space, without explicitly exploiting the structured spatial and temporal correlations that make precipitation fields highly compressible~\cite{fan2026physically}. Training and posterior sampling therefore remain computationally expensive, limiting applications to small spatial domains and modest temporal upscaling factors~\cite{schmidt2025generative}.

Several further limitations constrain the applicability of this framework to precipitation downscaling at the global scale. First, existing spatiotemporal posterior-sampling approaches have mainly been demonstrated in relatively homogeneous regions, where an unconditional generative prior can adequately represent the local distribution~\cite{glawion2025global,schmidt2025generative}. Extending this formulation to global precipitation is much more demanding, because a single unconditional prior must represent the full range of precipitation patterns across regions, seasons, and climate regimes. Second, previous work has typically used observation models designed for smoother variables such as temperature, pressure, and wind, for which linear spatial averaging with Gaussian likelihoods is often sufficient~\cite{schmidt2025generative}. Such models are not well suited to precipitation, whose coarse-scale values often reflect accumulated totals rather than instantaneous averages. Third, the likelihood score approximations used in posterior sampling can become numerically unstable at high noise levels, leading to unreliable gradient estimates and degraded sample quality~\cite{song2023solving,rozet2024learning}.

To address these challenges, we introduce Longwang, a zero-shot latent generative framework for joint spatiotemporal precipitation downscaling at the global scale. The name refers to the Dragon King, a deity in Chinese mythology who commands rain and storms. Longwang addresses the limitations identified above through the following contributions:

\begin{itemize}
    \item Longwang decouples generative prior learning from task-specific observation models, enabling zero-shot application to new downscaling tasks via Bayesian posterior sampling without retraining.

    \item The prior is trained in a compressed latent space, reducing memory and compute requirements to make global $\mathcal{O}(10\,\mathrm{km})$ downscaling with $32\times$ temporal refinement feasible on a single H100 80GB GPU.

    \item At inference, Longwang generates each ensemble member in approximately 30 seconds on the same hardware, making ensemble-based uncertainty quantification practical at operational scale.

    \item Longwang uses a physically informed observation operator that reflects precipitation accumulation across space and time, improving consistency with coarse-scale totals during posterior sampling.

    \item Longwang conditions the generative prior on spatiotemporal context, yielding a unified global downscaling framework that adapts to regional, seasonal, and climate regime variations in precipitation patterns.

    \item Longwang uses a stabilized posterior sampling scheme that improves numerical stability at large noise levels, while allowing the balance between the prior and likelihood to be controlled at inference.

\end{itemize}

\section{Results}
\label{sec:result}

\subsection{Reconstruction fidelity of the latent representation}
\label{sec:latent_fidelity}
Longwang learns its generative prior in the lower-dimensional latent space of a variational autoencoder (VAE) \cite{kingma2013auto}, which compresses precipitation fields by a factor of 4 along the temporal and spatial dimensions. We first assess the reconstruction skill of the VAE independently of the generative prior. ERA5 precipitation patches from the held-out test set are encoded into the latent space, decoded back to full resolution, and compared with the original fields.

Despite this compression, the VAE preserves the dominant statistical structure of precipitation fields (Supplementary Fig.~3). Spatially averaged monthly totals and lag-1 temporal autocorrelation closely match the reference data, with Pearson correlation coefficients of 0.999 for both metrics, indicating that mean rainfall and day-to-day persistence are retained. Reconstructed fields also remain highly correlated with the original fields at the grid-point level in both space and time. The daily precipitation intensity distribution is well reproduced, including the upper tail up to approximately 200 mm day$^{-1}$, although the 99.5th percentile is underestimated by approximately 3.8\% (1.2 mm day$^{-1}$). The primary degradation occurs at the finest spatial scales, where reconstructed fields exhibit reduced power at the highest spatial wavenumbers. This smoothing is expected when VAE reconstructions are obtained by decoding the posterior mean, which averages over plausible fine-scale structures~\cite{rombach2022high}. In the full Longwang pipeline, sampling from the diffusion prior in latent space partially restores this fine-scale variability.

\begin{figure}[tbp]
\centering
\includegraphics[width=\textwidth]{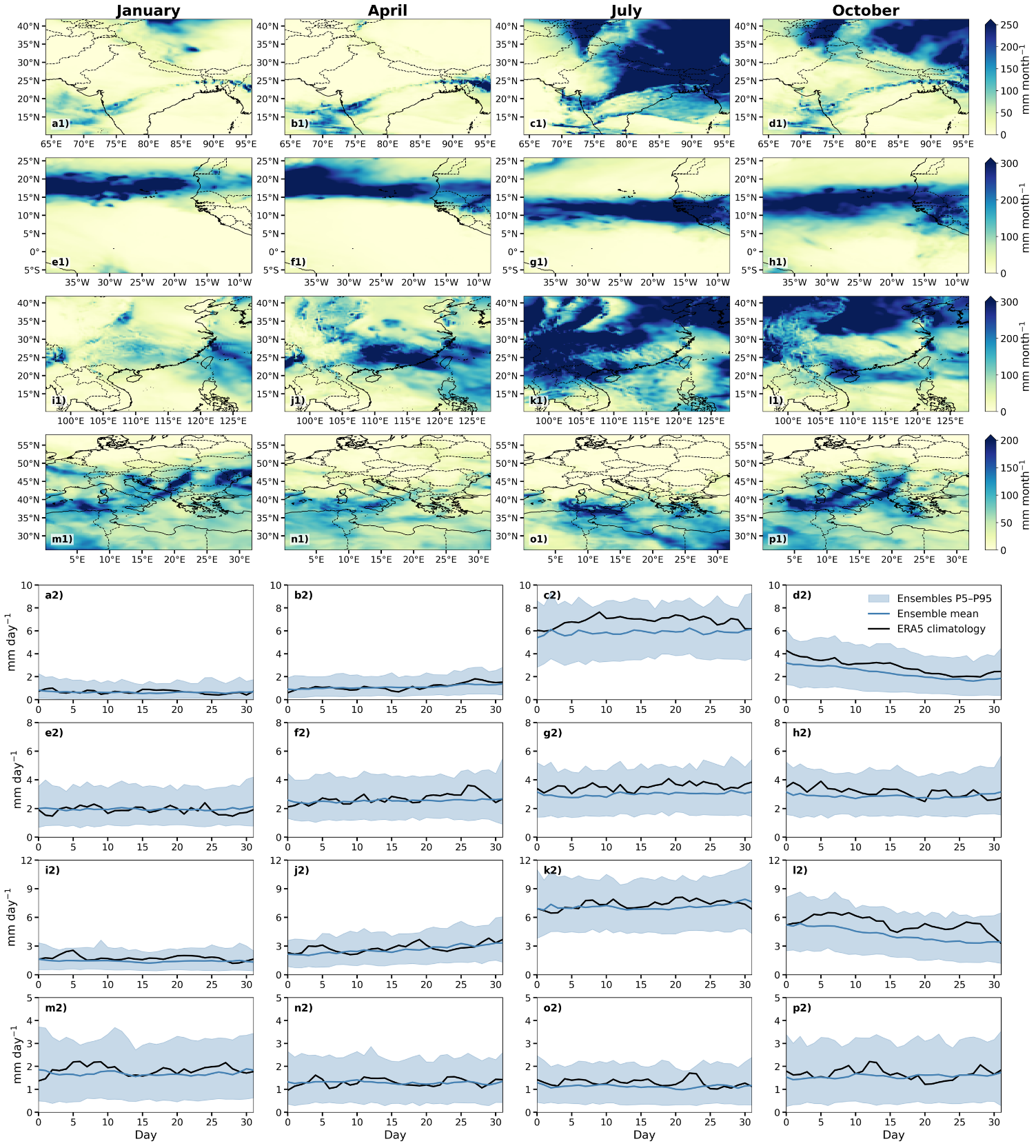}
\caption{\textbf{Precipitation patches sampled across seasons and regions from the conditional generative prior.} Samples are generated by the latent diffusion model and decoded to physical space using the VAE. Results are shown for four approximate geographic domains: India (\textbf{a--d}), Atlantic ITCZ (\textbf{e--h}), East China (\textbf{i--l}) and Western Europe (\textbf{m--p}). Columns correspond to January, April, July and October. Panels \textbf{a1--p1} show ensemble-mean monthly total precipitation computed from 50 samples. Panels \textbf{a2--p2} show the corresponding spatially averaged daily precipitation evolution for each region. The black line denotes the ERA5 climatology computed from the test set. The blue line denotes the generated ensemble mean, and shading indicates the 5th--95th percentile range across the same 50 generated samples.}
\label{fig:cond_prior_samples}
\end{figure}

\subsection{Regional and seasonal structure in the conditional prior}
\label{sec:cond_prior}

We next evaluate the conditional generative prior in Longwang. Unlike standard posterior sampling approaches that use an unconditional score-based diffusion prior~\cite{rozet2023score,schmidt2025generative}, Longwang conditions the prior on the calendar month and geographic center of the target patch, providing the model with seasonal and regional context during generation. To assess whether the prior learns to use these controls, we sample directly from it and compare the generated precipitation fields with ERA5 climatology. The evaluation covers four climatically distinct regions: India, the Atlantic Intertropical Convergence Zone (ITCZ), East China, and Western Europe. For each region, we consider four months spanning the seasonal cycle. For each region-month pair, we draw 50 samples and evaluate both the ensemble mean and the spread across realizations.

The ensemble mean reproduces the characteristic precipitation pattern of each region (Fig.~\ref{fig:cond_prior_samples} a1--p1). Over India, it captures the strong contrast between the July monsoon and the much drier conditions in January and April~\cite{turner2012climate} (Fig.~\ref{fig:cond_prior_samples} a1--d1). In the Atlantic ITCZ, it generates a zonal rainband that shifts meridionally with season, consistent with the observed migration of tropical convergence~\cite{schneider2014migrations} (Fig.~\ref{fig:cond_prior_samples} e1--h1). Over East China, it reproduces the summer wet season associated with the Meiyu front and the East Asian monsoon~\cite{ding2020multiscale} (Fig.~\ref{fig:cond_prior_samples} i1--l1), whereas over Western Europe it reproduces the comparatively weak seasonal cycle characteristic of the region~\cite{zveryaev2004seasonality} (Fig.~\ref{fig:cond_prior_samples} m1--p1).

The conditional prior also produces physically meaningful ensemble spread. For each region and month, the spatially averaged daily precipitation time series closely follows the corresponding ERA5 climatological reference, while the 5th--95th percentile envelope spans realistic day-to-day variability around the mean (Fig.~\ref{fig:cond_prior_samples} a2--p2). Spatial maps of the 5th and 95th percentiles further distinguish drier and wetter realizations while preserving coherent regional precipitation patterns (Supplementary Figs.~4 and 5). Together, these results show that Longwang uses spatiotemporal context to represent region- and season-specific precipitation climatology and variability, rather than collapsing to a generic precipitation pattern.

\subsection{Spatiotemporal downscaling of ERA5 reanalysis}
\label{sec:downscaling}

We evaluate Longwang in a zero-shot downscaling setting using the held-out ERA5 test set. Starting from each spatiotemporal ERA5 test sample, consisting of daily $0.25^\circ$ precipitation fields over a 32-day window, we apply the physically informed accumulation operator (Section~\ref{subsec:operator}) to obtain the corresponding monthly total precipitation field at $2^\circ$ resolution. Longwang is then tasked with reconstructing the original daily $0.25^\circ$ precipitation sequence from this coarse monthly input. This corresponds to an $8\times$ increase in spatial resolution and a $32\times$ increase in temporal resolution. The evaluation set contains 28 test patches spanning diverse precipitation regimes, including tropical convergence zones, major monsoon systems, midlatitude storm tracks, and subtropical and tropical continental regions (Supplementary Table~1). Performance is measured using complementary metrics (Supplementary Section~5) that assess spatial fidelity, marginal distribution agreement, probabilistic calibration, physical conservation, precipitation intermittency, extreme-value behavior, and temporal coherence. 

We compare Longwang with two baselines. The first is an unconditional generative prior baseline trained under the same setup as Longwang, so this comparison isolates the contribution of context conditioning. The second baseline is bilinear spatial interpolation with uniform temporal disaggregation (Bi+UT): a simple non-parametric reference that upsamples the $2^\circ$ monthly total to $0.25^\circ$ resolution and distributes the result uniformly across the 32-day window, yielding a daily precipitation field at the target spatiotemporal resolution.

The Bi+UT baseline performs well on the monthly total spatial reconstruction metric ($R^2=0.89$, Table~\ref{tab:evaluation}), because the low-resolution input is derived directly from the high-resolution field and does not require any bias correction. It also achieves nearly zero mass conservation error (MCE), as total precipitation is almost preserved by construction. These scores, however, reflect agreement in monthly aggregates rather than realistic reconstruction of daily precipitation fields. The limitation is evident in the Wasserstein-1 distance ($2.90$, compared with $0.56$ for Longwang; Table~\ref{tab:evaluation}) and in an R95p fraction of zero, indicating that none of its predictions exceed the ground-truth 95th-percentile wet-day threshold. This failure to generate extremes follows from the smoothing introduced by interpolation, while uniform temporal disaggregation spreads wet cells across all days, inflating the wet fraction and suppressing the dry-day fraction.

\begin{figure}[tbp]
\centering
\includegraphics[width=\textwidth]{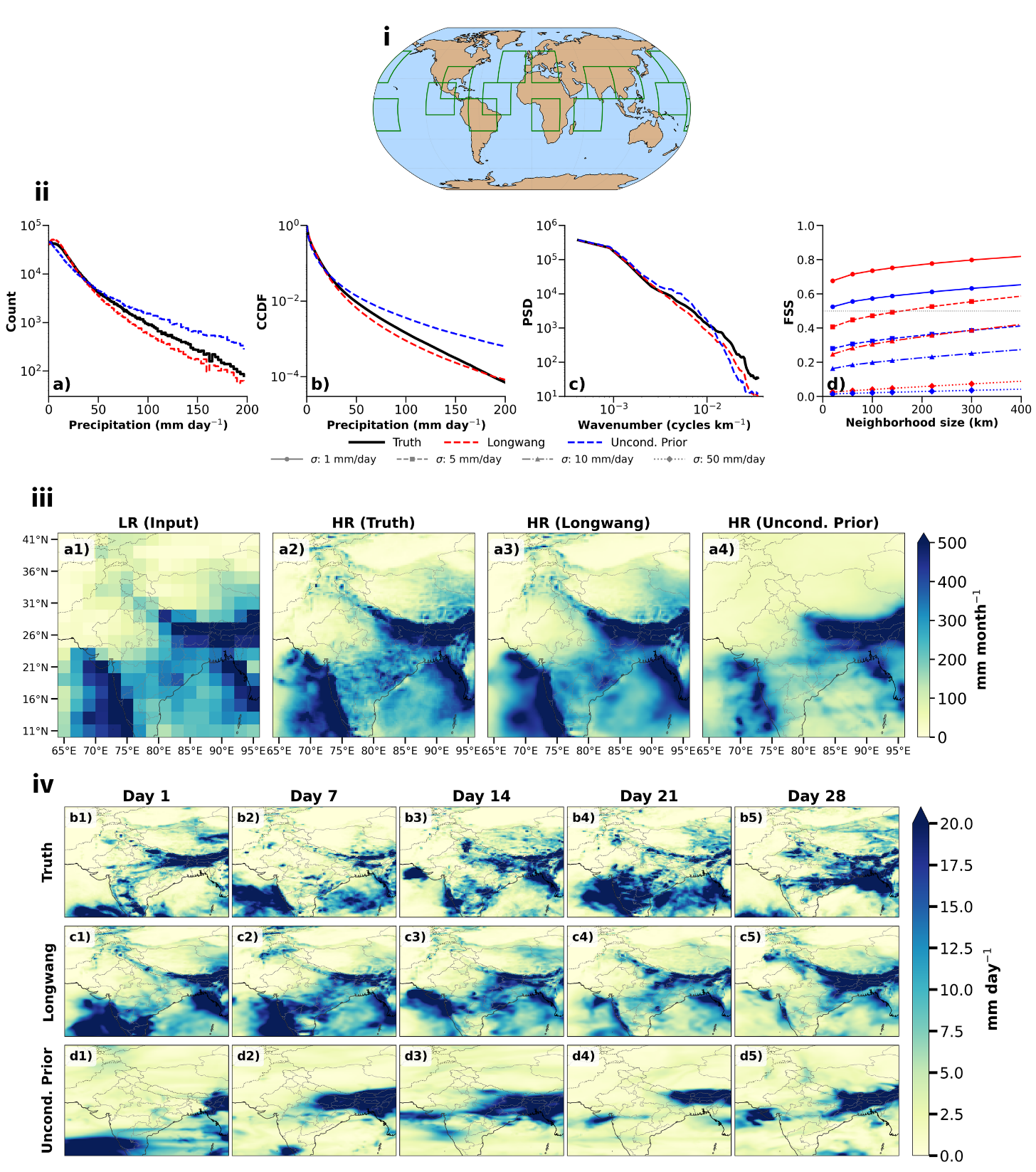}
\caption{\textbf{Comparison of Longwang with an unconditional generative prior on the held-out ERA5 test set.} \textbf{i}, Geographic distribution of the evaluated regions, indicated by green boxes. \textbf{ii}, Aggregate evaluation metrics across all test samples, as defined in Supplementary Information Section~5: \textbf{a}, histogram of daily precipitation; \textbf{b}, complementary cumulative distribution function (CCDF) of daily precipitation; \textbf{c}, isotropic power spectral density (PSD) of monthly precipitation totals; and \textbf{d}, fractions skill score (FSS) as a function of neighbourhood size for precipitation thresholds $\sigma \in \{1, 5, 10, 50\}$\,mm\,day$^{-1}$. \textbf{iii}, Monthly total precipitation for the Indian monsoon domain in July: \textbf{a1}, low-resolution input (LR, $2^\circ \times 2^\circ$); \textbf{a2}, high-resolution ground truth (HR, $0.25^\circ \times 0.25^\circ$); \textbf{a3}, Longwang; and \textbf{a4}, unconditional generative prior. \textbf{iv}, Daily precipitation snapshots for the same domain: \textbf{b1--b5}, ground truth; \textbf{c1--c5}, Longwang; and \textbf{d1--d5}, unconditional generative prior. Maps for Longwang and the unconditional prior baseline in \textbf{iii} and \textbf{iv} show ensemble means across posterior samples.}
\label{fig:test_era5}
\end{figure}

\begin{table*}[t]
\caption{Quantitative evaluation of bilinear spatial interpolation with uniform temporal disaggregation (Bi+UT), unconditional (Uncond) and conditional (Longwang) generative priors on the ERA5 test set. Results are reported as mean (5th, 95th percentile) over all test samples. Distributional statistics are reported for the ground truth (GT) and all three methods. See Supplementary Section~5 for metric definitions.}\label{tab:evaluation}
\small
\begin{tabular*}{\textwidth}{@{\extracolsep\fill}lcccc}
\toprule
\multicolumn{5}{@{}l@{}}{\textit{Performance Scores}} \\
\midrule
Model
  & $R^2$ $\uparrow$
  & Wasserstein (mm) $\downarrow$
  & CRPS (mm) $\downarrow$
  & MCE (\%) $\downarrow$ \\
\midrule
Bi+UT
  & $0.89$ (0.80, 0.95)
  & $2.90$ (1.40, 4.31)
  & ---
  & $0.01$ (0.00, 0.02) \\
Uncond
  & $0.63$ (0.20, 0.88)
  & $1.56$ (0.51, 3.39)
  & $2.89$ (1.38, 4.52)
  & $11.64$ (2.60, 28.01) \\
Longwang
  & $0.90$ (0.84, 0.95)
  & $0.56$ (0.22, 0.99)
  & $2.72$ (1.28, 4.26)
  & $3.66$ (1.26, 7.85) \\
\midrule
\multicolumn{5}{@{}l@{}}{\textit{Distributional Statistics}} \\
\midrule
Model
  & Wet Day Frac
  & Dry Day Frac
  & R95p Frac
  & Lag-1 Corr \\
\midrule
GT
  & 0.52 (0.28, 0.71)
  & 0.23 (0.07, 0.50)
  & 0.26 (0.20, 0.33)
  & 0.57 (0.28, 0.80) \\
Bi+UT
  & $0.80$ (0.55, 0.99)
  & $0.04$ (0.00, 0.20)
  & $0.00$ (0.00, 0.00)
  & --- \\
Uncond
  & $0.45$ (0.29, 0.63)
  & $0.27$ (0.11, 0.46)
  & $0.39$ (0.24, 0.57)
  & $0.47$ (0.33, 0.60) \\
Longwang
  & $0.56$ (0.32, 0.75)
  & $0.20$ (0.06, 0.44)
  & $0.21$ (0.15, 0.27)
  & $0.49$ (0.32, 0.66) \\
\botrule
\end{tabular*}
\end{table*}

Longwang outperforms the unconditional generative prior across all metrics (Table~\ref{tab:evaluation}). It achieves a higher mean $R^2$ ($0.90$ versus $0.63$) and a substantially higher 5th-percentile $R^2$ ($0.84$ versus $0.20$), indicating that the conditional prior is more robust on difficult samples. Because the two models share the same architecture, training data, likelihood, and posterior sampling procedure, this comparison isolates the effect of conditioning in the prior. The unconditional prior must represent global precipitation variability with a single distribution, whereas the conditional prior uses month and location to concentrate probability mass on patterns consistent with the target regime. This allows Longwang to retain a single unified global framework while adapting the prior to regional and seasonal specific precipitation patterns. Importantly, the improvement does not come at the cost of probabilistic calibration. The two models achieve similar continuous ranked probability scores (CRPS), suggesting that Longwang's gains are not due to overconfident predictions and reduced uncertainty spread.

Differences between the two priors are also evident in precipitation frequency and intensity. The wet-cell fraction measures how often precipitation occurs, whereas the R95p fraction measures the share of total rainfall contributed by extreme events. Relative to ERA5, the unconditional prior underestimates the wet-cell fraction and overestimates the R95p fraction, while Longwang remains closer to the reference for both metrics (Table~\ref{tab:evaluation}). These errors indicate that the unconditional prior tends to satisfy the coarse monthly accumulation constraint by concentrating rainfall into fewer, more intense cells, rather than distributing precipitation across the broader range of moderate intensities observed in ERA5. This behavior is also visible in the daily intensity distribution, where the unconditional prior overrepresents both near-dry and high-intensity cells (Fig.~\ref{fig:test_era5}ii,a,b).

The advantage of conditioning is not limited to marginal precipitation statistics. Aggregated across all test samples, the isotropic power spectra show that Longwang better preserves high-frequency signals than the unconditional prior (Fig.~\ref{fig:test_era5}ii,c). The fractions skill score (FSS), which evaluates spatial agreement after allowing for neighbourhood-scale displacement, is also consistently higher for Longwang across precipitation thresholds and neighbourhood sizes (Fig.~\ref{fig:test_era5}ii,d). These results indicate that conditioning improves not only aggregate reconstruction skill, but also the representation of fine-scale spatial structure. A representative July case over the Indian monsoon region illustrates how these aggregate differences appear in individual reconstructions (Fig.~\ref{fig:test_era5}iii,iv). In the monthly accumulated fields, Longwang recovers fine-scale precipitation structure more faithfully than the unconditional prior (Fig.~\ref{fig:test_era5}iii,a1--a4). The contrast is even clearer in the daily fields (Fig.~\ref{fig:test_era5}iv,b1--d5). Although reconstructing 32 daily maps from a single monthly-total input is highly ill posed, Longwang produces coherent fine-scale rainfall patterns that are more consistent with the ERA5 reference.

\subsection{Downscaling and bias correction of climate model outputs}
\label{sec:bias_correction}

\begin{figure}[tbp]
\centering
\includegraphics[width=\textwidth]{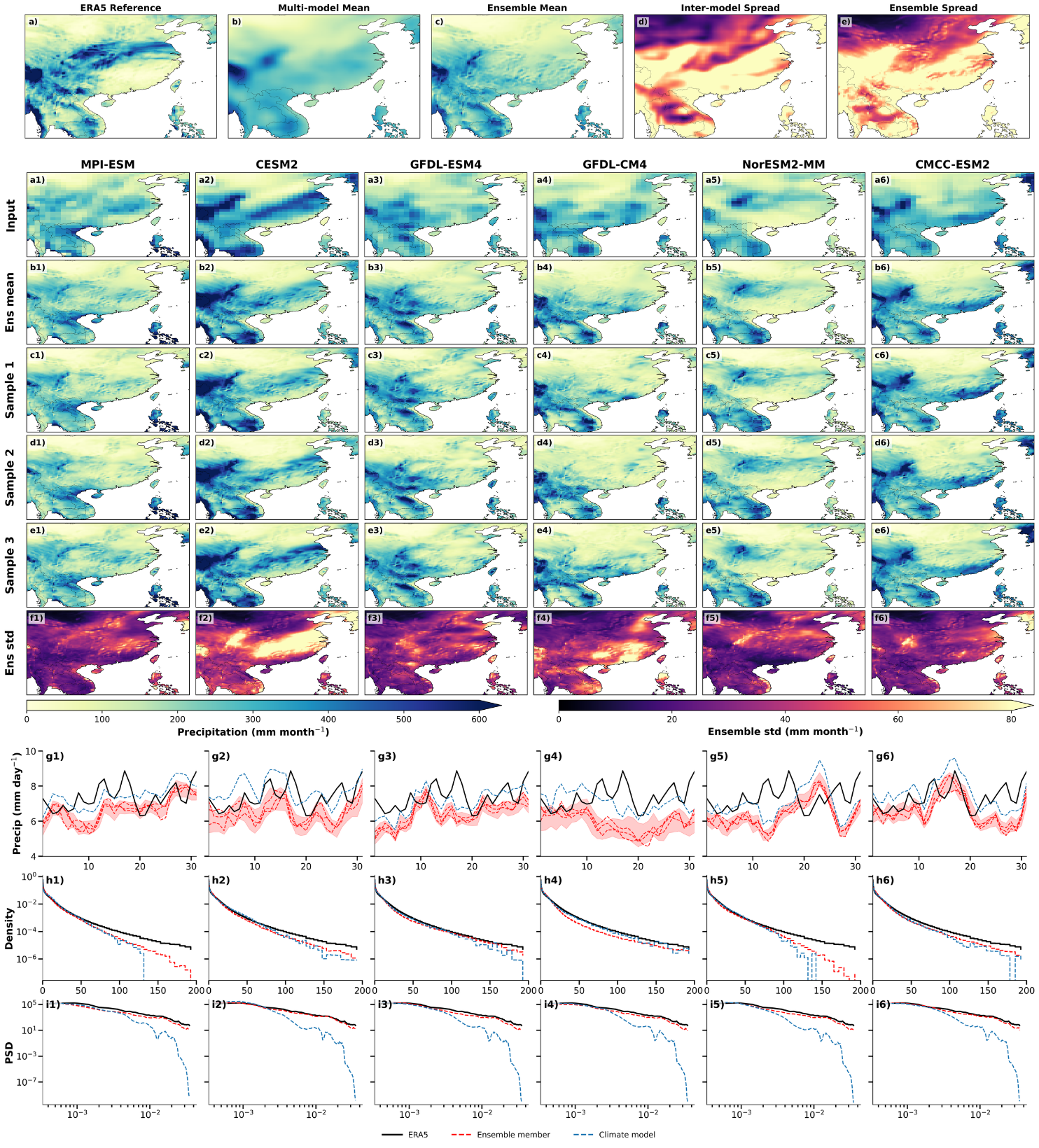}
\caption{\textbf{Stochastic downscaling of six CMIP6 climate models over East China for July 2005--2010.} \textbf{(a--e)} Monthly precipitation totals for July 2007: \textbf{(a)} ERA5 reference at $0.25^\circ$; \textbf{(b)} mean of the six climate model inputs after bilinear upsampling to the ERA5 grid; \textbf{(c)} mean of generated ensemble members by Longwang across the six models; \textbf{(d)} standard deviation across the six climate model inputs; \textbf{(e)} standard deviation across the generated ensemble members. \textbf{(a1--a6)} Coarse climate model inputs at approximately $1^\circ$ for July 2007. \textbf{(b1--b6)} Per-model ensemble means of the 50 Longwang downscaling samples. \textbf{(c1--c6, d1--d6, e1--e6)} Three individual ensemble members per model. \textbf{(f1--f6)} Per-model standard deviation across the 50 ensemble members. \textbf{(g1--g6)} Daily spatially averaged precipitation climatology for each model, averaged across the six July periods. The red shaded band shows the 5--95\% range across the ensemble members. \textbf{(h1--h6)} Probability density of precipitation values per model, pooled across all six years; for the ensemble, the curve is computed using one randomly selected member per year. \textbf{(i1--i6)} Isotropic spatial power spectral density of monthly precipitation fields per model, averaged over the six years; climate model inputs are bilinearly interpolated to the ERA5 grid.}
\label{fig:cmip_china_main}
\end{figure}

We next apply Longwang to historical precipitation simulations from six different climate models participating in the Coupled Model Intercomparison Project Phase 6 (CMIP6)~\cite{eyring2016overview}: MPI-ESM1-2-HR~\cite{muller2018higher}, CESM2~\cite{danabasoglu2020community}, GFDL-ESM4~\cite{dunne2020gfdl}, GFDL-CM4~\cite{held2019structure}, NorESM2-MM~\cite{seland2020overview}, and CMCC-ESM2~\cite{lovato2022cmip6}. Detailed descriptions of these datasets are provided in Section~\ref{subsec:data}. The analysis focuses on July precipitation over East China during 2005--2010. In this setting, the input consists of daily approximately $1^\circ$ precipitation fields, so the task is purely spatial downscaling to $0.25^\circ$ without temporal refinement. This provides a multi-model test of Longwang's flexibility across datasets and downscaling settings. We compare the downscaled fields with ERA5 over the same region and period. Since the simulations and reanalysis do not represent the same weather realizations, ERA5 is not used as ground truth. It instead provides a reference for comparing fine-scale spatial patterns and precipitation intensity distributions.

In this setting, we also test how posterior sampling can be adjusted at inference time by tuning the relative influence of the learned prior and the likelihood guidance. The observation operator enters through a likelihood guidance term that measures consistency with the coarse input (Fig.~\ref{fig:schematic}(iii)). This control is governed by two hyperparameters: $\tau_{\mathrm{start}}$, which determines when likelihood guidance begins during reverse diffusion, and $\sigma_y$, the assumed observation noise, which controls the strength of each likelihood gradient update (Supplementary Section~4). Increasing $\tau_{\mathrm{start}}$ introduces guidance earlier in the reverse trajectory, including during high-noise stages that affect large-scale structure, whereas decreasing $\sigma_y$ strengthens the pull toward the coarse input at each guided step.

Figure~\ref{fig:cmip_china_main} shows results under a relatively strong likelihood guidance ($\tau_{\mathrm{start}} = 0.5$, $\sigma_y = 15$). The downscaled ensemble mean for each model therefore preserves its own large-scale precipitation pattern (Fig.~\ref{fig:cmip_china_main} a1--a6 versus b1--b6). Longwang's job is to add ERA5-informed fine-scale spatial structure that is absent from the coarse inputs. A complementary case with a weaker likelihood guidance ($\tau_{\mathrm{start}} = 0.2$, $\sigma_y = 100$) is shown in Supplementary Fig.~6. In that setting, the prior learned from ERA5 has a strong influence on the large-scale patterns of the posterior samples. The ensemble mean of each downscaled climate model output therefore shows less diversity and aligns more closely with the ERA5 reference in some regions (Supplementary Fig.~6 a1--a6 versus b1--b6). In both settings, individual posterior samples look spatially coherent and physically plausible while showing clear diversity across realizations (Fig.~\ref{fig:cmip_china_main} c1--e6). Longwang therefore represents a distribution of plausible high-resolution precipitation fields rather than collapsing to the ensemble mean. 

The ensemble standard deviation reveals spatially varying uncertainty, with a larger spread in some regions than in others (Fig.~\ref{fig:cmip_china_main} f1--f6). This uncertainty also depends on the posterior sampling hyperparameters: reducing likelihood guidance increases the influence of the prior and leads to larger ensemble spread (Supplementary Fig.~6 f1--f6). The spatially averaged daily precipitation time series shows day-to-day variability across the July period (Fig.~\ref{fig:cmip_china_main} g1--g6), with a larger spread under weaker likelihood guidance. We also characterize how downscaling affects inter-model spread in addition to the per-model downscaling fidelity. The spread across Longwang's generated ensemble members is comparable in magnitude and spatial patterns to the inter-model spread among the six CMIP6 inputs (Fig.~\ref{fig:cmip_china_main} d,e). This suggests that stochastic downscaling introduces a substantial component of uncertainty, with spatial variability of a similar scale to the differences among climate models. However, these two spreads should not be interpreted as directly equivalent measures of uncertainty. The inter-model spread reflects differences among six climate models, whereas the Longwang ensemble spread includes both model-to-model differences and within-model stochastic variability from the generated samples.

For MPI-ESM and NorESM2-MM, the coarse simulations underrepresent the upper tail of the precipitation intensity distribution, with a cutoff near 140~mm. Longwang helps fill in this missing high-intensity range and therefore produces extreme precipitation values more closely aligned with ERA5 reference (Fig.~\ref{fig:cmip_china_main} h1,h5). This behavior reflects bias correction of high-intensity precipitation events that are underrepresented in some lower-resolution climate model simulations. The spatial power spectra further show that Longwang restores fine-scale variability missing from the coarse simulations, with the downscaled fields following ERA5 more closely at high wavenumbers (Fig.~\ref{fig:cmip_china_main} i1--i6).

\subsection{Generalization to future climate projections}
\label{sec:climate_projection}

We further apply Longwang to future climate projections from MPI-ESM. To assess generalization under substantial distribution shift, we consider three scenarios: the moderate-emissions pathway SSP2-4.5, the high-emissions pathway SSP5-8.5, and the SRM scenario G6sulfur, in which stratospheric aerosol injection is applied on top of SSP5-8.5 forcing to partially offset greenhouse-gas warming and keep global mean temperature closer to an SSP2-4.5 trajectory~\cite{kravitz2015geoengineering,visioni2021identifying}. All three scenarios lie outside the historical period used to train Longwang, and G6sulfur additionally introduces a sustained forcing pathway through continuous stratospheric aerosol injection, with no close analogue in the training data. The three scenarios are known to produce distinct regional precipitation responses, particularly over the West African monsoon region, where geoengineering can weaken Sahel rainfall through displacement of the intertropical convergence zone (ITCZ)~\cite{bonou2023stratospheric}. We therefore focus our evaluation on this region.

\begin{figure}[tbp]
\centering
\includegraphics[width=0.95\textwidth]{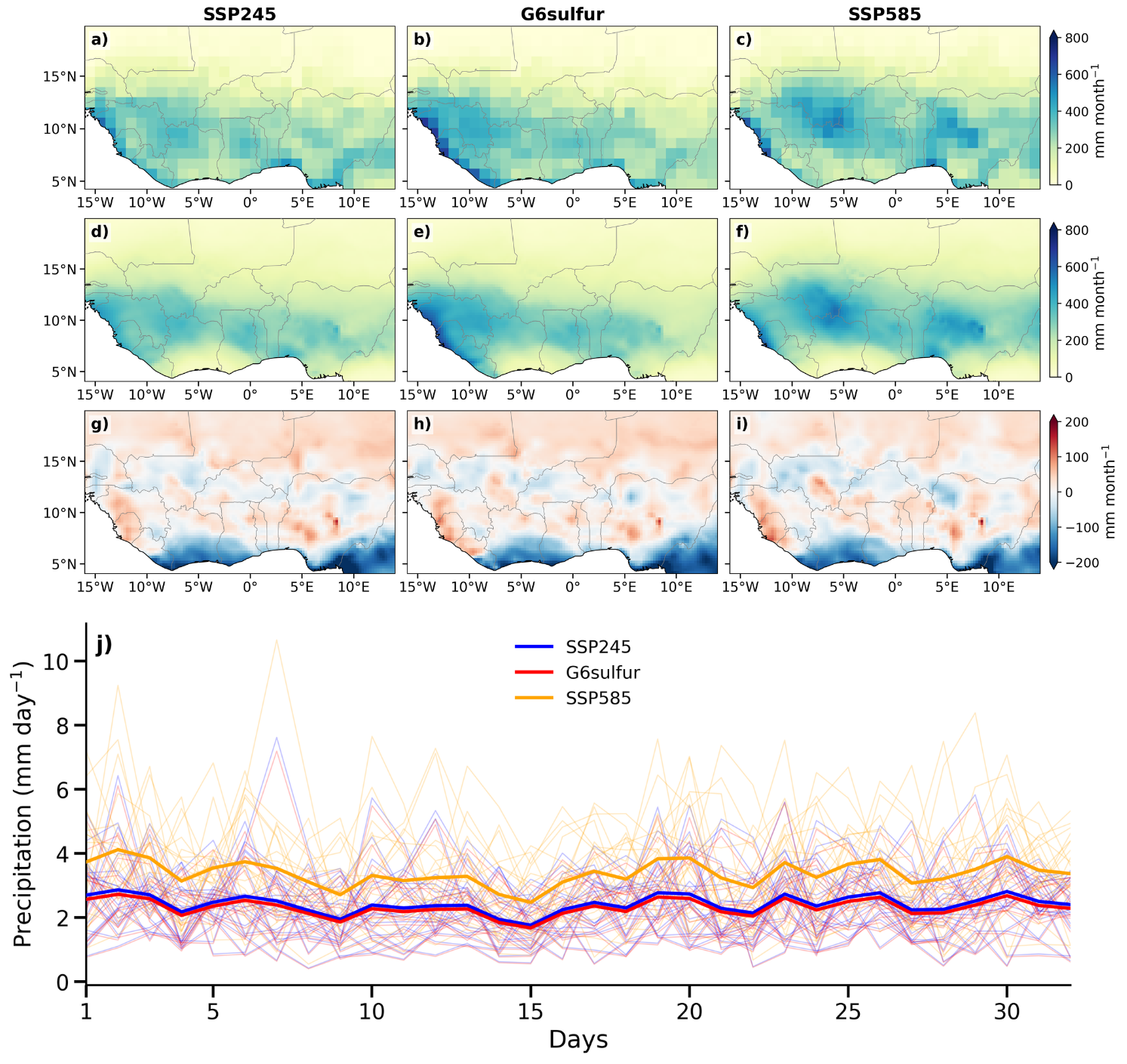}
\caption{\textbf{Application of Longwang to spatiotemporal downscaling of climate projections under the SSP2-4.5, G6sulfur, and SSP5-8.5 scenarios over West Africa in August 2041.} \textbf{a--c}, Monthly precipitation from MPI-ESM-1-2-HR simulations at approximately $1^\circ$ resolution for \textbf{a}, SSP2-4.5; \textbf{b}, G6sulfur; and \textbf{c}, SSP5-8.5. \textbf{d--f}, Monthly precipitation at $0.25^\circ$ resolution downscaled by Longwang (20-member ensemble mean), using the MPI-ESM-1-2-HR simulations in \textbf{a--c} as inputs. \textbf{g--i}, Differences between the downscaled fields in \textbf{d--f} and the corresponding bilinearly interpolated coarse inputs from \textbf{a--c}. \textbf{j}, Spatially averaged daily precipitation time series over the northern Sahel ($14^\circ$--$18^\circ$N, $5^\circ$W--$10^\circ$E) for SSP2-4.5 (blue), G6sulfur (red), and SSP5-8.5 (orange). Light lines show the 20 individual ensemble members per scenario; solid bold lines show the ensemble mean.}
\label{fig:g6sulfur_ssp}
\end{figure}

Across all scenarios, Longwang generates spatially coherent high-resolution precipitation fields from the corresponding MPI-ESM inputs while preserving the scenario-specific large-scale precipitation patterns (Fig.~\ref{fig:g6sulfur_ssp} a--f). Here, we use a relatively strong likelihood guidance with $\tau_{\mathrm{start}} = 0.4$ and $\sigma_y = 40$. To isolate the fine-scale structure added by Longwang, we compare the downscaled fields with the linearly upsampled input fields (Fig.~\ref{fig:g6sulfur_ssp} g--i). The added fine-scale patterns are broadly similar across G6sulfur, SSP5-8.5, and SSP2-4.5, suggesting that the ERA5-trained prior mainly contributes realistic small-scale features that look similar across scenarios. Beyond spatial refinement, Longwang also performs temporal downscaling by mapping monthly inputs to daily high-resolution precipitation sequences. Figure~\ref{fig:g6sulfur_ssp}j shows the spatially averaged daily precipitation time series over the northern Sahel for all generated ensemble members. The G6sulfur trajectory remains below SSP5-8.5 and close to SSP2-4.5 across most ensemble members throughout August, consistent with a sustained weakening of West African monsoon precipitation under stratospheric aerosol injection~\cite{bonou2023stratospheric}. This indicates that Longwang preserves the precipitation contrast among scenarios while generating plausible daily variability at high spatial and temporal resolution.

\section{Discussion}
\label{sec:discussion}
We introduced Longwang as a flexible framework for zero-shot, stochastic spatiotemporal downscaling of precipitation at the global scale. Rather than learning a diffusion prior directly over high-resolution precipitation fields, we first train a VAE to compress precipitation patches into a lower-dimensional latent space and learn the diffusion prior there. At the patch sizes considered here ($32 \times 128 \times 128$), this latent-space formulation makes training and inference tractable under our compute budget, whereas diffusion in physical space would be prohibitive. Because the prior is learned independently of the observation model, we cast downscaling as a Bayesian inverse problem solved via posterior sampling, which allows the same prior to be reused across downscaling tasks without retraining. 

With a conditional diffusion prior, Longwang yields better posterior sampling quality for downscaling reanalysis data than an otherwise identical model using an unconditional prior. Whether an unconditional prior suffices in this framework depends on both the breadth of the learned distribution and the informativeness of the likelihood. When the prior is trained over a relatively homogeneous domain, its probability mass concentrates around a narrow range of precipitation regimes; when the downscaling factor is modest, the likelihood term carries substantial information about the target high-resolution field. Under such conditions, an unconditional prior can remain effective \cite{schmidt2025generative}. In our setting, however, the spatiotemporal patches are cropped globally across regions and seasons, so a single unconditional prior would struggle to represent the full range of precipitation regimes within one marginal distribution. Meanwhile, the inverse problem considered here is severely ill-posed: with 8-fold spatial and 32-fold temporal downscaling, the coarse input constrains large-scale monthly accumulation but provides little signal about day-to-day variability or fine-scale spatial structure. Posterior sampling therefore requires a more informative prior to resolve the missing degrees of freedom, which is what conditioning provides.

Beyond reanalysis, Longwang can be applied to jointly bias correct and downscale climate model outputs. It can recover the tail of the precipitation intensity distribution and restore the spatial power spectrum at high wavenumbers missed by coarse numerical simulations. This framework can be further applied to different future climate projection scenarios under substantial distribution shift. On SSP2-4.5, SSP5-8.5 and the G6sulfur stratospheric aerosol injection scenarios, Longwang adds plausible fine-scale spatial structure while preserving the large-scale patterns of each scenario. Although the prior is trained on present day statistics, the likelihood signal anchors the posterior to the coarse projection input, so the forcing response is not collapsed onto present day climatology. This suggests Longwang has the potential to generalize to a wide range of climate projections and scenarios where high-resolution output would otherwise be prohibitively expensive to simulate directly.

The decoupling between prior and likelihood in Longwang provides more than zero-shot flexibility. It also allows for the relative contributions of the learned generative prior and likelihood guidance via tunable hyperparameters during inference. When the low-resolution input is reliable, applying likelihood guidance at higher noise levels and with stronger gradient updates keeps the posterior tightly anchored to the coarse field. The prior hence mainly contributes to fine-scale variability. When the input is some coarse numerical simulation with known biases, weakening the likelihood guidance gives the prior more influence. This tunability is what enables Longwang to support both super-resolution-style reconstruction and bias correction within a single unified framework.

Several limitations and extensions remain. First, the likelihood operator used here treats the low-resolution input as an idealized degradation of the high-resolution field. However, real low-resolution fields sometimes also contain physical biases that cannot be captured by accumulation from high-resolution fields. Extending the likelihood to account for such biases is a natural extension. Second, our conditioning choice was deliberately minimal to avoid extrapolation during inference. Conditioning on the full date, for example, would make applications to future climate projections depend on years unseen during training, creating unwanted extrapolation. In contrast, conditioning on patch location allows generalization to new centers through spatial interpolation. Future work could incorporate richer conditioning information, such as indicators of dominant climate modes El Niño--Southern Oscillation (ENSO)~\cite{zhao2024explainable}. Third, the present study focuses on daily precipitation. Extending the framework to hourly downscaling from daily inputs, should require only appropriate training data and observation operators. Another related direction is to broaden the training data used to learn the prior. Longwang currently learns its prior from ERA5 reanalysis, but future work could instead use physically consistent high-resolution climate simulations, such as variable-resolution global models with regional refinement to approximately $0.1^\circ$~\cite{xu2021evaluating}, to better align the learned prior with climate dynamics. Beyond precipitation, extending Longwang to multivariable downscaling of fields such as temperature and surface radiation would broaden its use in climate impact applications. This multivariable setting may be challenging, however, because variables with different spectral structures and degrees of intermittency may not share a clean common latent representation.

\section{Methods}
\label{sec:methods}

\subsection{Data}
\label{subsec:data}

We use daily total precipitation from the ERA5 reanalysis~\cite{hersbach2020era5} at $0.25^{\circ}$ (${\sim}25\,\text{km}$) spatial resolution over the global domain spanning 1979--2024. We begin in 1979 because the modern satellite era substantially improves the availability of global atmospheric observations assimilated into reanalysis products. The data are split chronologically into 1979--2014 for training, 2015--2019 for validation, and 2020--2024 for testing. Precipitation is highly skewed, with values spanning several orders of magnitude. To compress its dynamic range, we apply a shifted logarithmic transform:
\begin{equation}\label{eq:log_transform}
    \tilde{x} = \log(x_{\mathrm{mm}} + \varepsilon) - \log(\varepsilon), \quad \varepsilon = 10^{-4},
\end{equation}
where $x_{\mathrm{mm}}$ is precipitation in millimeters. The transformed values are then normalized to $[-1, 1]$ using the global minimum and maximum computed from the training set. We tile the global domain into spatiotemporal patches of shape $32 \times 128 \times 128$ (time $\times$ latitude $\times$ longitude). Each patch covers 32~days and $32^{\circ} \times 32^{\circ}$ in space. Patches are extracted using a sliding window with 50\% overlap in both the temporal and spatial dimensions. For each patch, we also record the center latitude, center longitude, and calendar month at the temporal midpoint of the window. These variables provide the conditioning context used to train Longwang (Section~\ref{subsec:diffusion}).

For evaluation, we consider three data settings. First, we evaluate spatiotemporal downscaling performance on the held-out ERA5 test split. The coarse input is constructed from the high-resolution target using the accumulation operator described in Section~\ref{subsec:operator}. This setting provides paired high- and low-resolution fields and allows direct quantitative evaluation against the ground truth. Second, we evaluate joint downscaling and bias correction on historical precipitation simulations from six different global climate models participating in the Coupled Model Intercomparison Project Phase 6 (CMIP6)~\cite{eyring2016overview}. These include the Max Planck Institute Earth System Model version 1.2 in its high-resolution configuration (MPI-ESM1-2-HR)~\cite{muller2018higher}; the Community Earth System Model version 2 (CESM2)~\cite{danabasoglu2020community}; the Geophysical Fluid Dynamics Laboratory Earth System Model version 4 (GFDL-ESM4)~\cite{dunne2020gfdl}; the Geophysical Fluid Dynamics Laboratory Coupled Model version 4 (GFDL-CM4)~\cite{held2019structure}; the Norwegian Earth System Model version 2 in its medium-resolution configuration (NorESM2-MM)~\cite{seland2020overview}; and the Centro Euro-Mediterraneo sui Cambiamenti Climatici Earth System Model version 2 (CMCC-ESM2)~\cite{lovato2022cmip6}. For this historical multi-model evaluation, we use daily precipitation over East China during July 2005--2010. The model outputs are treated as coarse inputs and downscaled to the ERA5 grid with Longwang. As the climate simulations and ERA5 reanalysis do not represent the same realized weather events, ERA5 is used as a reference for fine-scale spatial structure and precipitation intensity distributions rather than as the ground truth. Third, we test generalization under distribution shift using future climate projections from MPI-ESM1-2-HR over West Africa. This experiment uses three future forcing scenarios: the moderate-emissions Shared Socioeconomic Pathway SSP2-4.5, the high-emissions Shared Socioeconomic Pathway SSP5-8.5, and the Geoengineering Model Intercomparison Project G6sulfur scenario~\cite{kravitz2015geoengineering}. 

\subsection{Evaluation metrics}
\label{subsec:eval_metrics}
Evaluating precipitation downscaling requires more than pointwise accuracy: realistic precipitation fields should also capture fine-scale spatial structure, the tails of intensity distributions, and temporal variability. We therefore use complementary metrics to assess different aspects of the generated precipitation ensembles.

In the paired ERA5 test setting, where high-resolution ERA5 fields provide corresponding ground truth, we use the coefficient of determination ($R^2$) of monthly accumulated precipitation to evaluate large-scale spatial pattern recovery, Wasserstein-1 distance to compare marginal precipitation intensity distributions, and the continuous ranked probability score (CRPS) to assess probabilistic calibration of the generated ensemble. We also define mass conservation error (MCE) to quantify bias in domain-integrated precipitation. To evaluate precipitation occurrence and extremes, we compute wet-day and dry-day fractions, together with the normalized R95p fraction, which measures the contribution of heavy precipitation to total rainfall. Temporal consistency is assessed using the lag-1 autocorrelation of spatially averaged precipitation time series. Spatial organization of precipitation events is evaluated using the fractions skill score (FSS), while power spectral density (PSD) is used to assess whether generated fields reproduce the scale-dependent spatial variability. We also examine complementary cumulative distribution functions (CCDFs) to compare tail behavior of daily precipitation intensities.

For CMIP6 historical and future climate experiments, the climate model inputs and ERA5 reference fields are not paired realizations of the same weather events. Evaluation in these settings therefore emphasizes statistical realism rather than pointwise reconstruction. We therefore focus on whether the downscaled products reproduce ERA5-like fine-scale spatial patterns, precipitation intensity distributions, and spectral power across spatial scales. Full mathematical definitions and implementation details for all metrics mentioned above are provided in the Supplementary Section 5.

\subsection{Variational autoencoder}
\label{subsec:vae}
We use a variational autoencoder (VAE)~\cite{kingma2013auto} to learn a lower-dimensional latent representation of precipitation patches and train the diffusion model in this latent space. The VAE downsamples each spatial and temporal axis by a factor of 4 and expands the channel dimension from 1 to 16, mapping an input $\mathbf{x} \in \mathbb{R}^{1 \times 32 \times 128 \times 128}$ to a latent $\mathbf{z} \in \mathbb{R}^{16 \times 8 \times 32 \times 32}$. The encoder parameterises a diagonal Gaussian posterior $q(\mathbf{z}\mid\mathbf{x})$ through its mean and log-variance, and $\mathbf{z}$ is sampled using the reparameterisation trick~\cite{kingma2013auto}. The VAE is trained with the loss
\begin{equation}\label{eq:vae_loss}
    \mathcal{L}_{\text{VAE}} = \lambda_{\text{rec}}\,\mathcal{L}_{1} + \lambda_{\text{sg}}\,\mathcal{L}_{\nabla_{xy}} + \lambda_{\text{tg}}\,\mathcal{L}_{\nabla_{t}} + \lambda_{\text{KL}}\,D_{\text{KL}}\!\left(q(\mathbf{z}|\mathbf{x}) \,\|\, p(\mathbf{z})\right),
\end{equation}
where $\mathcal{L}_{1}$ is the pointwise L1 reconstruction loss, $\mathcal{L}_{\nabla_{xy}}$ and $\mathcal{L}_{\nabla_{t}}$ are spatial and temporal gradient losses computed with central finite differences, and $D_{\text{KL}}$ is the Kullback--Leibler (KL) divergence between the encoder posterior and a standard Gaussian prior. By penalizing errors in spatial and temporal derivatives, the gradient losses effectively upweight high-frequency components that the L1 term alone tends to smooth~\cite{lyu2024multi}. The KL weight $\lambda_{\text{KL}}$ is kept small because the downstream diffusion model does not require a perfectly Gaussian latent space. A light KL penalty is sufficient to regularize the latent distribution while preserving reconstruction quality~\cite{rombach2022high}. Further details of the VAE architecture and training procedure are provided in Supplementary Section~1.

\subsection{Score-based diffusion model}
\label{subsec:diffusion}
The score-based diffusion model~\cite{song2020score} is trained in the latent space to define a generative prior over compressed precipitation patches. We consider two variants of this prior. The unconditional prior learns a single distribution across all patch locations and calendar months. The conditional prior instead uses the patch center latitude, sine--cosine encodings of the center longitude, and sine--cosine encodings of the calendar month to represent region- and season-specific precipitation patterns. These conditioning variables are encoded with a small multilayer perceptron (MLP) and injected into every residual block through adaptive group normalization~\cite{dhariwal2021diffusion}.

The stochastic differential equation (SDE) formulation provides access to a time-dependent score function for the latent prior. At inference, this score can be combined with an observation likelihood to perform zero-shot posterior sampling under new downscaling configurations, without retraining the prior (Section~\ref{subsec:posterior}). We adopt the variance-preserving (VP) SDE, whose forward process progressively corrupts a clean latent sample $\mathbf{z} \sim p(\mathbf{z})$ via
\begin{equation}\label{eq:forward_sde}
    \mathrm{d}\mathbf{z}(\tau) = f(\tau)\,\mathbf{z}(\tau)\,\mathrm{d}\tau + g(\tau)\,\mathrm{d}\mathbf{w}(\tau),
\end{equation}
where $f(\tau)$ and $g(\tau)$ are scalar drift and diffusion coefficients, $\mathbf{w}(\tau)$ is a standard Wiener process, and $\tau \in [0,1]$ indexes the noise level. Because the SDE is linear, the perturbation kernel has a closed-form Gaussian distribution:
\begin{equation}\label{eq:perturbation_kernel}
    p_\tau(\mathbf{z}(\tau) \mid \mathbf{z}_0) = \mathcal{N}\bigl(\mu(\tau)\,\mathbf{z}_0,\; \sigma(\tau)^2\,\mathbf{I}\bigr),
\end{equation}
where $\mu(\tau)$ and $\sigma(\tau)$ are the signal and noise coefficients given by the cosine schedule~\cite{nichol2021improved}. This closed-form kernel allows us to train the model by directly sampling noisy latents, rather than simulating the forward SDE. We train a network $\boldsymbol{\varepsilon}_\theta(\mathbf{z}(\tau), \tau, \mathbf{c})$ to predict the Gaussian noise $\boldsymbol{\varepsilon} \sim \mathcal{N}(\mathbf{0}, \mathbf{I})$ added during the forward process, using the denoising score matching (DSM) objective
\begin{equation}\label{eq:dsm_loss}
    \mathcal{L}_{\text{DSM}} = \mathbb{E}_{p(\mathbf{z})\,\mathcal{U}(\tau)\,\mathcal{N}(\boldsymbol{\varepsilon})}\Bigl[\bigl\lVert\boldsymbol{\varepsilon}_\theta\!\bigl(\mu(\tau)\,\mathbf{z} + \sigma(\tau)\,\boldsymbol{\varepsilon},\;\tau,\;\mathbf{c}\bigr) - \boldsymbol{\varepsilon}\bigr\rVert_2^2\Bigr],
\end{equation}
where $\tau \sim \mathcal{U}(0,1)$ is sampled uniformly over the diffusion time interval, and $\mathbf{c}$ is the conditioning vector. The predicted noise defines the score through
\begin{equation}\label{eq:score_from_eps}
    \nabla_{\mathbf{z}(\tau)} \log p_\tau\!\bigl(\mathbf{z}(\tau)\mid \mathbf{c}\bigr)
    \approx
    -\frac{1}{\sigma(\tau)}\,\boldsymbol{\varepsilon}_\theta(\mathbf{z}(\tau), \tau, \mathbf{c}).
\end{equation}
At inference, this learned score is used in the reverse SDE~\cite{anderson1982reverse},
\begin{equation}\label{eq:reverse_sde}
    \mathrm{d}\mathbf{z}(\tau) = \Bigl[f(\tau)\,\mathbf{z}(\tau) - g(\tau)^2 \,\nabla_{\mathbf{z}(\tau)} \log p_\tau\!\bigl(\mathbf{z}(\tau) \mid \mathbf{c}\bigr)\Bigr]\mathrm{d}\tau + g(\tau)\,\mathrm{d}\bar{\mathbf{w}}(\tau),
\end{equation}
which is simulated backward from $\tau = 1$ to $\tau = 0$ to transform Gaussian noise into samples from the learned latent prior. Further details on the diffusion architecture and training procedure are provided in Supplementary Section~2.

\subsection{Physically informed observation operator}
\label{subsec:operator}
Posterior sampling (Section~\ref{subsec:posterior}) requires an observation operator $\mathcal{A}$ that maps a high-resolution daily precipitation field to the coarse-resolution observation $\mathbf{y}$. Prior work on smoother variables has often used linear spatial averaging~\cite{schmidt2025generative}, but coarse-scale precipitation is typically reported as an accumulated total rather than an instantaneous average. We therefore define $\mathcal{A}$ as an area-weighted spatial average followed by temporal accumulation. Let $x(t, \phi, \lambda)$ denote the high-resolution daily precipitation field, where $\phi$ and $\lambda$ are latitude and longitude. Each coarse cell $(i,j)$ corresponds to a spherical region $\Omega_{ij} \subset S^2$, and observations are aggregated over a $T$-day window $[t_0, t_0 + T]$. The surface-area element on the sphere is $dA = \cos\phi\, d\phi\, d\lambda$, and the area of a coarse cell is $|\Omega_{ij}| = \int_{\Omega_{ij}} dA$. The observation operator is therefore
\begin{equation}\label{eq:operator_continuous}
    (\mathcal{A}x)(\Omega_{ij}, t_0) \;=\; \int_{t_0}^{t_0+T} \frac{1}{|\Omega_{ij}|}\int_{\Omega_{ij}} x(t, \phi, \lambda)\, dA\, dt.
\end{equation}
The inner spatial integral computes the area-weighted average of the daily field over the coarse cell, and the outer temporal integral accumulates these daily precipitation amounts over the $T$-day window. The composite operator $\mathcal{A}$ is linear, differentiable, and preserves area-weighted totals, properties that are essential for gradient-based posterior sampling.

\subsection{Zero-shot posterior sampling}
\label{subsec:posterior}

\subsubsection*{Posterior formulation}
We formulate precipitation downscaling as a Bayesian inverse problem in the latent space. Let $\mathbf{x} \in \mathbb{R}^{T \times H \times W}$ denote a high-resolution daily precipitation field and let $\mathbf{y}$ denote a coarse-resolution, temporally aggregated observation. The forward degradation model is
\begin{equation}\label{eq:degradation}
    \mathbf{y} = \mathcal{A}(\mathbf{x}) + \boldsymbol{\eta}, \qquad \boldsymbol{\eta} \sim \mathcal{N}(\mathbf{0}, \sigma_y^2 \mathbf{I}),
\end{equation}
where $\mathcal{A}$ is the operator defined in Section~\ref{subsec:operator} and $\sigma_y$ absorbs the mismatch between the idealised operator and the true coarse--fine relationship, which in practice also reflects model biases and other discrepancies. Because the diffusion prior is defined over latent variables $\mathbf{z}$ rather than over $\mathbf{x}$, the posterior is hence formulated in latent space via Bayes' rule:
\begin{equation}\label{eq:posterior}
    p(\mathbf{z} \mid \mathbf{y}, \mathbf{c}) \;\propto\; p(\mathbf{z} \mid \mathbf{c})\; p\!\bigl(\mathbf{y} \mid \mathcal{A}(\mathcal{D}(\mathbf{z}))\bigr),
\end{equation}
where $\mathcal{D}$ is the VAE decoder and $\mathbf{c}$ denotes the spatiotemporal conditioning vector that is fed to the conditional prior ($\mathbf{y}$ is assumed to be conditionally independent of $\mathbf{c}$ given $\mathbf{z}$). Sampling from this posterior requires the score $\nabla_{\mathbf{z}(\tau)} \log p(\mathbf{z}(\tau) \mid \mathbf{y}, \mathbf{c})$, which decomposes as
\begin{equation}\label{eq:score_decomp}
    \nabla_{\mathbf{z}(\tau)} \log p(\mathbf{z}(\tau) \mid \mathbf{y}, \mathbf{c})
    =
    \nabla_{\mathbf{z}(\tau)} \log p_\tau(\mathbf{z}(\tau) \mid \mathbf{c})
    +
    \nabla_{\mathbf{z}(\tau)} \log p(\mathbf{y} \mid \mathbf{z}(\tau)).
\end{equation}
The prior score is provided by the trained network $\boldsymbol{\varepsilon}_\theta$ (Section~\ref{subsec:diffusion}). To compute the likelihood score, we first form a Tweedie-type estimate of the clean latent variable at each diffusion step,
\begin{equation}\label{eq:tweedie}
    \hat{\mathbf{z}}_0(\mathbf{z}(\tau)) = \frac{\mathbf{z}(\tau) - \sigma(\tau)\,\boldsymbol{\varepsilon}_\theta(\mathbf{z}(\tau), \tau, \mathbf{c})}{\mu(\tau)},
\end{equation}
then decode this estimate via VAE and apply the observation operator to obtain a predicted coarse observation $\hat{\mathbf{y}} = \mathcal{A}(\mathcal{D}(\hat{\mathbf{z}}_0))$. Further details on posterior sampling implementation are provided in Supplementary Section~4.

\subsubsection*{Guidance control}
Applying likelihood guidance at every diffusion step is numerically unstable. At high noise levels, the Tweedie estimate $\hat{\mathbf{z}}_0$ becomes unreliable because the factor $1/\mu(\tau)$ amplifies estimation error~\cite{song2023solving}. In addition, the Gaussian likelihood $\mathcal{N}(\mathbf{y} \mid \hat{\mathbf{y}}, \sigma_y^2 \mathbf{I})$ used in standard diffusion posterior sampling is overconfident, as it ignores the substantial posterior uncertainty in the clean latent given $\mathbf{z}(\tau)$~\cite{chung2022diffusion,rozet2024learning}.

We therefore use a noise-aware posterior-guidance scheme with two modifications. First, likelihood guidance is activated only below a noise threshold $\tau_{\mathrm{start}}$. For $\tau \geq \tau_{\mathrm{start}}$, sampling follows the prior alone, allowing the diffusion process to first form a globally plausible latent structure before the coarse observations are introduced. Second, we inflate the likelihood variance with a noise-dependent term:
\begin{equation}\label{eq:approx_likelihood}
    p(\mathbf{y} \mid \mathbf{z}(\tau)) \;\approx\; \mathcal{N}\!\Bigl(\mathbf{y}\;\Big|\;\hat{\mathbf{y}},\;\sigma_y^2\,\mathbf{I} + \gamma \bigl(\sigma(\tau)/\mu(\tau)\bigr)^{2} \mathbf{I}\Bigr),
\end{equation}
where $\gamma > 0$ controls the strength of the variance inflation. This corresponds to a Gaussian approximation in which the conditional distribution $p(\mathbf{z}_0 \mid \mathbf{z}(\tau))$ is treated as having nonzero covariance $\mathbb{V}[\mathbf{z}_0 \mid \mathbf{z}(\tau)]$~\cite{rozet2024learning}, rather than as a point mass at $\hat{\mathbf{z}}_0$ as in standard diffusion posterior sampling. The term $\gamma(\sigma(\tau)/\mu(\tau))^2 \mathbf{I}$ provides a simple scalar approximation to this covariance, becoming large at high noise levels and vanishing as $\tau \to 0$. The resulting posterior-guided noise prediction is
\begin{equation}\label{eq:guided_eps}
    \tilde{\boldsymbol{\varepsilon}}_\theta(\mathbf{z}(\tau), \tau \mid \mathbf{y}, \mathbf{c}) = \boldsymbol{\varepsilon}_\theta(\mathbf{z}(\tau), \tau, \mathbf{c}) \;-\; \sigma(\tau)\,\nabla_{\mathbf{z}(\tau)} \log p(\mathbf{y} \mid \mathbf{z}(\tau)),
\end{equation}
which replaces the prior noise prediction in the reverse SDE (Eq.~\eqref{eq:reverse_sde}). Because this posterior-guidance scheme is not tied to a specific observation operator $\mathcal{A}$, the same trained prior can be applied zero-shot to any downscaling task with a differentiable forward operator. In practice, the relative influence of the likelihood is controlled by the posterior-sampling hyperparameters: $\tau_{\mathrm{start}}$ determines when guidance is activated, while $\sigma_y$ controls how strongly the likelihood gradient pulls the sample toward the coarse observation once guidance is applied.

\section{Data availability}
ERA5 reanalysis data are available from the Copernicus Climate Data Store~\cite{hersbach2023era5single}. The CMIP6 historical simulations and future projections are publicly available through the Earth System Grid Federation (ESGF)~\cite{cinquini2014esgf} at \url{https://esgf.github.io/}.

\backmatter

\bmhead{Acknowledgements}
Support for D.V. and Y.M. was provided by the Quadrature Climate Foundation. 

\bmhead{Author contributions}
Y.W. conceived the study, developed the method, performed the experiments, analyzed the results and wrote the manuscript. D.V. supervised the project and contributed to the study design, interpretation of results and manuscript writing. All authors reviewed and approved the manuscript.

\bmhead{Competing interests}
The authors declare no competing interests.

\bmhead{Supplementary information}
Supplementary information is available for this paper. Correspondence and requests for materials should be addressed to Y.W.

\bibliography{reference}

\end{document}


\title{Supplementary Information for Longwang}
\author{Yue Wang \and Daniele Visioni}
\maketitle

\tableofcontents
\clearpage


\clearpage
\section{VAE Architecture and Training}\label{sec:supp_vae}
\begin{figure}[h]
\centering
\includegraphics[width=0.85\textwidth]{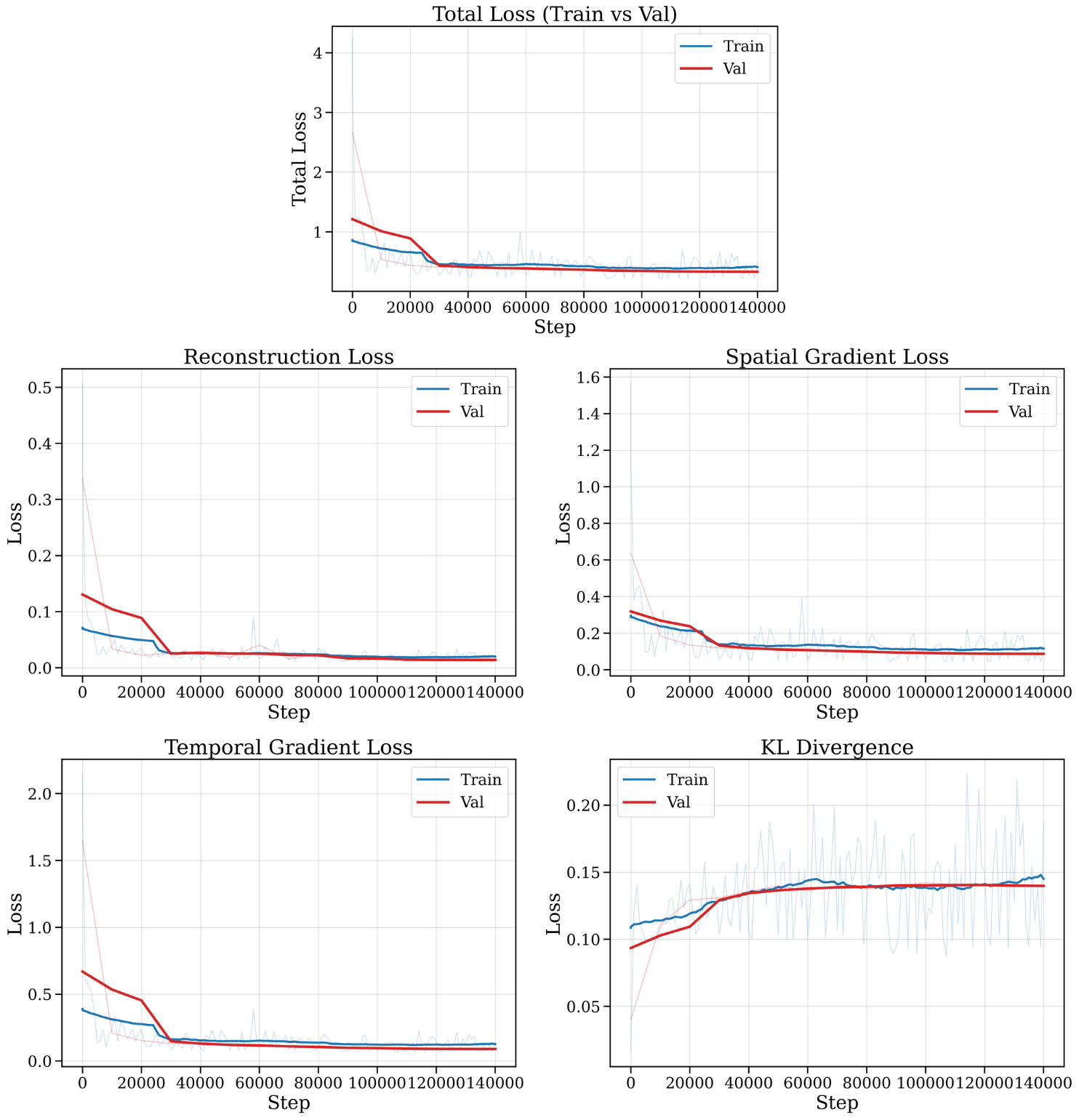}
\caption{Training and validation losses for the VAE.}
\label{fig:vae_loss}
\end{figure}

The encoder of the variational autoencoder (VAE)~\cite{kingma2013auto} consists of three resolution levels with channel multipliers $1, 2, 4$ relative to a base channel width of 128, yielding feature maps with 128, 256, and 512 channels. Each level contains four residual blocks with group normalization~\cite{wu2018group} and SiLU activations~\cite{elfwing2018sigmoid}. The decoder mirrors the encoder structure, using transposed convolutions for upsampling. The model has approximately 200 million parameters. The VAE is trained with the Adam optimizer~\cite{kingma2014adam} and a cosine learning rate schedule~\cite{loshchilov2016sgdr}, with a linear warm-up over the first 60k steps followed by decay from a peak learning rate of $2 \times 10^{-4}$ to $2 \times 10^{-5}$. Training is distributed across 8 NVIDIA H100 GPUs with a per-GPU batch size of 1 (effective batch size 8) for 140k steps. An exponential moving average (EMA)~\cite{morales2024exponential} of the model weights is maintained with a decay rate of 0.999. The loss weights in Equation~(2) of the main text are $\lambda_{\mathrm{rec}} = 1.0$, $\lambda_{\nabla_{xy}} = 10.0$, $\lambda_{\nabla_t} = 10.0$, and $\lambda_{\mathrm{KL}} = 10^{-6}$. The training and validation loss curves are shown in Figure~\ref{fig:vae_loss}.

\clearpage
\section{Diffusion Model Architecture and Training}\label{sec:supp_diffusion}
\begin{figure}[h]
\centering
\includegraphics[width=0.85\textwidth]{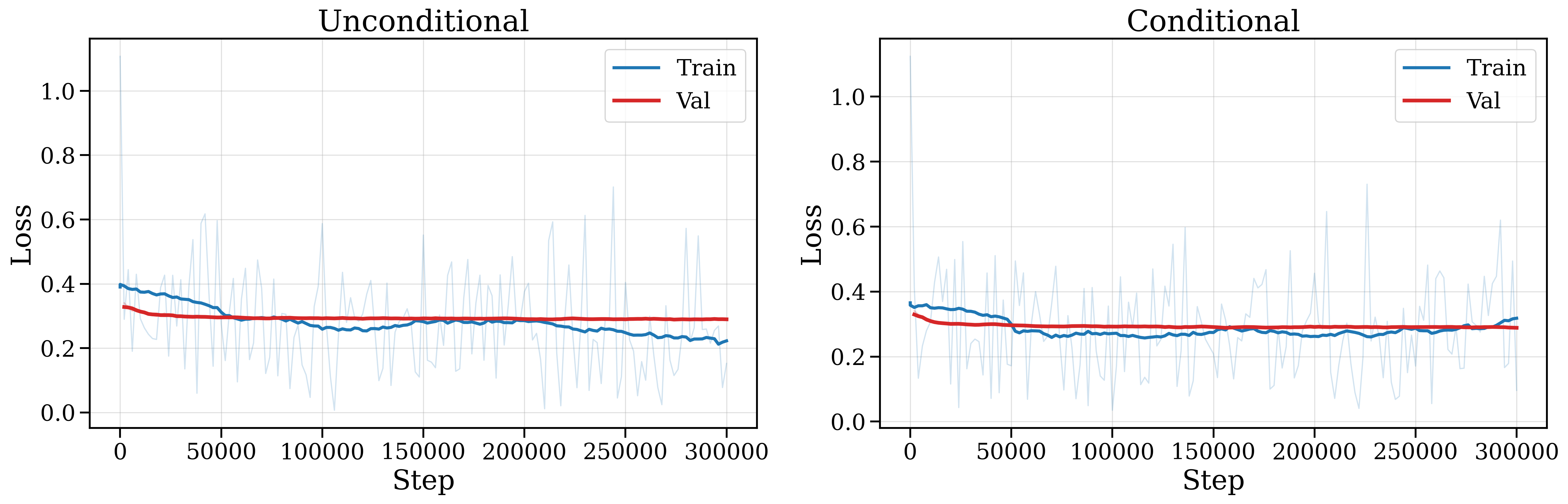}
\caption{Training and validation losses for the unconditional and conditional diffusion models.}
\label{fig:diffusion_loss}
\end{figure}

The noise prediction network $\boldsymbol{\varepsilon}_\theta$ is a 3D U-Net \cite{ronneberger2015u} with four resolution levels, a uniform channel width of 512, and four residual blocks per level. Multi-head self-attention \cite{vaswani2017attention} with 8 heads is applied at all three encoder and decoder levels and at the bottleneck. Diffusion timesteps are embedded via a sinusoidal positional encoding projected to 256 dimensions. The conditioning vector is mapped through a three-layer multilayer perceptron (MLP) \cite{gardner1998artificial} to the same 256-dimensional space and added to the timestep embedding before injection into each residual block via adaptive group normalization \cite{dhariwal2021diffusion}. The model has approximately 400 million parameters. The diffusion model is trained with the Adam optimizer, using a linear warm-up over 10k steps to a peak learning rate of $2 \times 10^{-4}$. Training is distributed across 8 NVIDIA H100 GPUs with a per-GPU batch size of 4 (effective batch size 32) for 300k steps. EMA is maintained with a decay rate of 0.9999. The variance-preserving SDE uses a cosine noise schedule with $\eta = 10^{-3}$. Before training, all latent channels are z-score normalized to zero mean and unit variance using statistics computed over the training set. The training and validation loss curves are shown in Figure~\ref{fig:diffusion_loss}.
\clearpage

\clearpage
\section{Reverse SDE Discretisation}\label{sec:supp_discretisation}
We derive the exponential integrator update rule~\cite{zhang2022fast} used to simulate the reverse SDE during inference. Starting from the probability flow ODE associated with the reverse SDE (Equation (7) in the main text), we drop the stochastic term to obtain the deterministic sampler:
\begin{equation}\label{eq:ode}
    \frac{\mathrm{d}\mathbf{z}}{\mathrm{d}\tau} = \frac{\mathrm{d}\log\mu}{\mathrm{d}\tau}\,\mathbf{z}(\tau) + \frac{g(\tau)^2}{2\,\sigma(\tau)}\,\boldsymbol{\varepsilon}_\theta(\mathbf{z}(\tau), \tau, \mathbf{c}),
\end{equation}
where we have substituted the score--noise relation $\nabla_{\mathbf{z}(\tau)}\log p_\tau(\mathbf{z}(\tau)\mid\mathbf{c}) = -\boldsymbol{\varepsilon}_\theta / \sigma(\tau)$. The first term is linear in $\mathbf{z}$, which admits an exact solution via the integrating factor $1/\mu(\tau)$. Defining $\mathbf{u}(\tau) = \mathbf{z}(\tau)/\mu(\tau)$, we obtain
\begin{equation}\label{eq:u_ode}
    \frac{\mathrm{d}\mathbf{u}}{\mathrm{d}\tau} = \frac{g(\tau)^2}{2\,\mu(\tau)\,\sigma(\tau)}\,\boldsymbol{\varepsilon}_\theta(\mathbf{z}(\tau), \tau, \mathbf{c}).
\end{equation}
Discretising from $\tau$ to $\tau - \Delta\tau$ and assuming $\boldsymbol{\varepsilon}_\theta$ is approximately constant over one step gives
\begin{equation}
    \mathbf{u}(\tau - \Delta\tau) \approx \mathbf{u}(\tau) + \boldsymbol{\varepsilon}_\theta \int_{\tau}^{\tau - \Delta\tau} \frac{g(s)^2}{2\,\mu(s)\,\sigma(s)}\,\mathrm{d}s.
\end{equation}
Substituting back $\mathbf{z} = \mu\,\mathbf{u}$ yields
\begin{equation}
    \mathbf{z}(\tau - \Delta\tau) = \frac{\mu(\tau - \Delta\tau)}{\mu(\tau)}\,\mathbf{z}(\tau) + \mu(\tau - \Delta\tau) \left[\int_{\tau}^{\tau - \Delta\tau} \frac{g(s)^2}{2\,\mu(s)\,\sigma(s)}\,\mathrm{d}s\right] \boldsymbol{\varepsilon}_\theta.
\end{equation}
For the variance-preserving SDE, the integral evaluates to $\sigma(\tau - \Delta\tau)/\mu(\tau - \Delta\tau) - \sigma(\tau)/\mu(\tau)$, which, after multiplying by $\mu(\tau - \Delta\tau)$, gives the coefficient $\sigma(\tau - \Delta\tau) - r\,\sigma(\tau)$ with $r \equiv \mu(\tau - \Delta\tau)/\mu(\tau)$. The final update rule is
\begin{equation}\label{eq:exp_integrator}
    \mathbf{z}(\tau - \Delta\tau) = r\,\mathbf{z}(\tau) + \bigl(\sigma(\tau - \Delta\tau) - r\,\sigma(\tau)\bigr)\,\boldsymbol{\varepsilon}_\theta(\mathbf{z}(\tau), \tau, \mathbf{c}).
\end{equation}
For posterior sampling, the prior noise prediction $\boldsymbol{\varepsilon}_\theta$ is replaced by the guided prediction $\tilde{\boldsymbol{\varepsilon}}_\theta$ from Equation (15) in the main text, and the update rule is otherwise unchanged.
\clearpage

\clearpage
\section{Posterior Sampling Implementation}
\label{sec:supp_sampling}
This section provides implementation details for the posterior sampling scheme described in Section~4.5 of the main text. At each reverse diffusion step, the latent variable is updated using a posterior-guided noise prediction. Likelihood guidance is activated only below a noise threshold $\tau_{\mathrm{start}}$. For $\tau \geq \tau_{\mathrm{start}}$, sampling follows the prior alone:
\begin{equation}\label{eq:supp_unguided_eps}
    \tilde{\boldsymbol{\varepsilon}}_\theta(\mathbf{z}(\tau), \tau \mid \mathbf{y}, \mathbf{c})
    =
    \boldsymbol{\varepsilon}_\theta(\mathbf{z}(\tau), \tau, \mathbf{c}).
\end{equation}
For $\tau < \tau_{\mathrm{start}}$, we apply likelihood guidance through
\begin{equation}\label{eq:supp_guided_eps}
    \tilde{\boldsymbol{\varepsilon}}_\theta(\mathbf{z}(\tau), \tau \mid \mathbf{y}, \mathbf{c})
    =
    \boldsymbol{\varepsilon}_\theta(\mathbf{z}(\tau), \tau, \mathbf{c})
    -
    \sigma(\tau)\,\nabla_{\mathbf{z}(\tau)} \log p(\mathbf{y}\mid \mathbf{z}(\tau)).
\end{equation}
The likelihood is approximated as
\begin{equation}\label{eq:supp_likelihood}
    p(\mathbf{y}\mid \mathbf{z}(\tau))
    \approx
    \mathcal{N}\!\Bigl(
    \mathbf{y}\;\Big|\;
    \hat{\mathbf{y}},
    \;
    \sigma_y^2 \mathbf{I}
    +
    \gamma \bigl(\sigma(\tau)/\mu(\tau)\bigr)^2 \mathbf{I}
    \Bigr),
\end{equation}
where $\sigma_y$ is the observation noise standard deviation and $\gamma$ controls the strength of the noise-dependent variance inflation. The term $\gamma(\sigma(\tau)/\mu(\tau))^2 \mathbf{I}$ serves as a scalar approximation to the posterior covariance of the clean latent given $\mathbf{z}(\tau)$. To evaluate the likelihood gradient $\nabla_{\mathbf{z}(\tau)}\log p(\mathbf{y} \mid \mathbf{z}(\tau))$, we first form the Tweedie estimate of the clean latent,
\begin{equation}\label{eq:supp_tweedie}
    \hat{\mathbf{z}}_0
    =
    \frac{\mathbf{z}(\tau)-\sigma(\tau)\,\boldsymbol{\varepsilon}_\theta(\mathbf{z}(\tau),\tau,\mathbf{c})}{\mu(\tau)},
\end{equation}
then decode it and apply the observation operator to obtain the predicted coarse observation,
\begin{equation}\label{eq:supp_yhat}
    \hat{\mathbf{y}}=\mathcal{A}(\mathcal{D}(\hat{\mathbf{z}}_0)).
\end{equation}
The gradient is computed by automatic differentiation through the chain
\[
\mathbf{z}(\tau)
\;\to\;
\hat{\mathbf{z}}_0
\;\to\;
\mathcal{D}(\hat{\mathbf{z}}_0)
\;\to\;
\hat{\mathbf{y}}
\;\to\;
\log p(\mathbf{y}\mid \mathbf{z}(\tau)).
\]
For downscaling the ERA5 test set, we use $\tau_{\mathrm{start}} = 0.5$, $\gamma = 10^{-2}$, $\sigma_y = 15$, and $N = 128$ reverse diffusion steps. Under this configuration, the full sampling procedure takes approximately 30 seconds per test sample per ensemble member on a single NVIDIA H100 GPU. For joint downscaling and bias correction of CMIP6 climate models, we evaluate two posterior-sampling settings with different strengths of likelihood guidance. The first uses $\tau_{\mathrm{start}} = 0.5$ and $\sigma_y = 15$, corresponding to relatively strong likelihood guidance and a weaker influence of the prior. The second uses $\tau_{\mathrm{start}} = 0.2$ and $\sigma_y = 100$, corresponding to weaker likelihood guidance and a stronger influence of the prior.
\clearpage

\clearpage
\section{Evaluation Metrics}
\label{sec:eval_metrics}
Let $X(t, \mathbf{x})$ denote the ground-truth precipitation field at day $t$ and spatial location $\mathbf{x} = (h, w)$, and let $\hat{X}(t, \mathbf{x})$ denote a single random sample draw from the predictive distribution. We use $\langle \cdot \rangle$ to denote averaging over the relevant indices (spatial, temporal, or ensemble). All metrics are computed per draw and averaged across the ensemble, except for the continuous ranked probability score \cite{hersbach2000decomposition}.

The coefficient of determination $R^2$ measures how well the spatial pattern of monthly accumulated precipitation is recovered. Let $S(\mathbf{x})$ and $\hat{S}(\mathbf{x})$ denote the monthly totals of truth and prediction. Then
\begin{equation}
    R^2 = 1 - \frac{\langle (S - \hat{S})^2 \rangle_{\mathbf{x}}}{\langle (S - \langle S \rangle_{\mathbf{x}})^2 \rangle_{\mathbf{x}}}.
\end{equation}

Wasserstein-1 distance compares the marginal distributions of precipitation intensities. Let $F$ and $\hat{F}$ denote the cumulative distribution functions of $X$ and $\hat{X}$. Then
\begin{equation}
    \mathcal{W} = \int_{0}^{\infty} \left| \hat{F}(x) - F(x) \right| dx.
\end{equation}

The continuous ranked probability score (CRPS) evaluates probabilistic calibration by jointly rewarding forecast accuracy and appropriate ensemble spread. At each spatiotemporal point, CRPS compares the predictive CDF $\hat{F}_{t, \mathbf{x}}$ to the observation $X(t, \mathbf{x})$:
\begin{equation}
    \mathrm{CRPS}(\hat{F}_{t, \mathbf{x}}, X(t, \mathbf{x})) = \int_{0}^{\infty} \bigl( \hat{F}_{t, \mathbf{x}}(x) - \mathbf{1}[x \geq X(t, \mathbf{x})] \bigr)^2 dx.
\end{equation}

The mass conservation error (MCE) quantifies how well total precipitation is conserved relative to the ground truth:
\begin{equation}
    \mathrm{MCE} = \frac{\bigl| \langle \hat{X} \rangle - \langle X \rangle \bigr|}{\langle X \rangle} \times 100\%,
\end{equation}
where $\langle \cdot \rangle$ denotes the spatiotemporal average over $(t, \mathbf{x})$.

Wet and dry day fractions characterize precipitation frequency statistics. A grid cell is classified as wet if its precipitation exceeds $1$\,mm\,day$^{-1}$ and as dry if it falls below $0.1$\,mm\,day$^{-1}$:
\begin{equation}
    f^{\mathrm{wet}} = \mathbb{P}(\hat{X} \geq 1), \qquad
    f^{\mathrm{dry}} = \mathbb{P}(\hat{X} < 0.1),
\end{equation}
estimated as empirical frequencies over all grid cells.

The R95p fraction quantifies how well a model represents the tail of the precipitation distribution. We adopt a normalised variant of the ETCCDI R95p index~\citep{zhang2011indices}: the fraction of total precipitation contributed by cells exceeding the $95$th percentile of the ground-truth wet-day distribution. The shared climatological threshold is
\begin{equation}
    Q_{95} = P_{95}\!\left( X \mid X \geq 1 \right),
\end{equation}
where $P_{95}$ denotes the 95th percentile. The R95p fraction for a field $Z \in \{X, \hat{X}\}$ is
\begin{equation}
    R^{95}(Z) = \frac{\langle Z \cdot \mathbf{1}[Z > Q_{95}] \rangle}{\langle Z \rangle}.
\end{equation}
Sharing $Q_{95}$ between truth and predictions evaluates both against the same physical reference point, while normalising each by its own total decouples tail shape from total-mass bias.

Lag-1 autocorrelation captures day-to-day temporal persistence of the domain-mean precipitation. Let $z(t) = \langle \hat{X}(t, \mathbf{x}) \rangle_{\mathbf{x}}$ denote the spatial mean at day $t$. Then
\begin{equation}
    \hat{\rho} = \frac{\mathrm{Cov}(z_t, z_{t+1})}{\mathrm{Var}(z_t)},
\end{equation}
where covariance and variance are estimated empirically across $t$.

The Fractions Skill Score (FSS) \cite{roberts2008scale} evaluates spatial co-location of precipitation events at multiple scales. For a precipitation threshold $\sigma$ and a uniform $n \times n$ box filter $\mathcal{B}_n$, define the smoothed neighborhood fractions of threshold exceedance:
\begin{equation}
    O^{(n)}(\sigma) = \mathcal{B}_n \!\left[ \mathbf{1}[X \geq \sigma] \right], \qquad
    F^{(n)}(\sigma) = \mathcal{B}_n \!\left[ \mathbf{1}[\hat{X} \geq \sigma] \right].
\end{equation}
The FSS at scale $n$ and threshold $\sigma$ is
\begin{equation}
    \mathrm{FSS}^{(n)}(\sigma) = 1 - \frac{\langle (O^{(n)} - F^{(n)})^2 \rangle}{\langle (O^{(n)})^2 \rangle + \langle (F^{(n)})^2 \rangle},
\end{equation}
where $\langle \cdot \rangle$ denotes spatial averaging. We compute FSS for each threshold $\sigma \in \{1, 5, 10, 50\}$\,mm\,day$^{-1}$.

The complementary cumulative distribution function (CCDF) of precipitation intensity quantifies the heavy-tailed marginal behavior of daily rainfall:
\begin{equation}
    \overline{F}(x) = \mathbb{P}(\hat{X} > x), \qquad
    \overline{F}_{\mathrm{gt}}(x) = \mathbb{P}(X > x),
\end{equation}
representing empirical exceedance probabilities at intensity $x$. We report $\overline{F}$ on doubly logarithmic axes, where heavy tails appear as power-law-like decays and discrepancies at extreme intensities are most visible.

The isotropic spatial power spectrum evaluates how precipitation variance is distributed across spatial scales. For a field $Z \in \{X,\hat{X}\}$, let $\mathcal{F}[Z_t](\mathbf{k})$ denote the two-dimensional Fourier transform of $Z(t,\mathbf{x})$. The isotropic spectrum is obtained by averaging Fourier power over circles of radius $\kappa$ in wavenumber space:
\begin{equation}
    E_Z(\kappa)
    =
    \frac{1}{|\mathbb{S}_{\kappa}|}
    \int_{\mathbb{S}_{\kappa}}
    \left|
    \mathcal{F}[Z_t](\mathbf{k})
    \right|^2
    dS(\mathbf{k}),
\end{equation}
where $\mathbb{S}_{\kappa}=\{\mathbf{k}:\|\mathbf{k}\|_2=\kappa\}$. We compare $\langle E_X(\kappa)\rangle_t$ and $\langle E_{\hat{X}}(\kappa)\rangle_t$ to assess whether generated fields reproduce the scale-dependent spatial variability of precipitation.

\clearpage
\section{Supplementary Tables}\label{sec:eval_regions}

\begin{table}[h]
\caption{Evaluation regions and months for posterior sampling. Each region is centred on a $32\degree \times 32\degree$ domain at 0.25\textdegree\ resolution.}\label{tab:eval_regions}%
\begin{tabular}{@{}lccll@{}}
\toprule
Region & Centre latitude & Centre longitude & Months & Regime \\
\midrule
\multicolumn{5}{@{}l}{\textit{Intertropical Convergence Zone}} \\
East Pacific         & 10\textdegree N  & 256\textdegree E & Jul, Aug     & ITCZ convection \\
Central Pacific      &  6\textdegree S  & 192\textdegree E & Jul, Aug     & ITCZ convection \\
West Pacific         & 10\textdegree N  & 160\textdegree E & Jan, Jul     & ITCZ convection \\
Atlantic             & 10\textdegree N  & 320\textdegree E & Jul, Aug     & ITCZ convection \\
Indian Ocean         &  6\textdegree S  &  64\textdegree E & Jan, Jul     & ITCZ convection \\
\addlinespace
\multicolumn{5}{@{}l}{\textit{Monsoon}} \\
India                & 26\textdegree N  &  80\textdegree E & Jul, Aug     & Indian summer monsoon \\
East Asia            & 26\textdegree N  & 112\textdegree E & Jun, Jul     & East Asian monsoon \\
Central Africa       &  6\textdegree S  &  16\textdegree E & Jan, Jul     & African monsoon \\
\addlinespace
\multicolumn{5}{@{}l}{\textit{Midlatitude storm tracks}} \\
North Atlantic       & 42\textdegree N  & 336\textdegree E & Jan, Dec     & Extratropical cyclones \\
North Pacific        & 42\textdegree N  & 176\textdegree E & Jan, Dec     & Extratropical cyclones \\
European Westerlies  & 42\textdegree N  &  16\textdegree E & Jan, Dec     & Frontal precipitation \\
\addlinespace
\multicolumn{5}{@{}l}{\textit{Subtropical and tropical continental}} \\
Amazon               &  6\textdegree S  & 304\textdegree E & Jan, Jul     & Tropical convection \\
Southeast US         & 26\textdegree N  & 272\textdegree E & Jul, Aug     & Subtropical convection \\
Southeast China      & 26\textdegree N  & 112\textdegree E & Apr, May     & Pre-monsoon (Meiyu) \\
\botrule
\end{tabular}
\footnotetext[1]{East Asia and Southeast China share the same grid centre (26\textdegree N, 112\textdegree E) but target distinct months: June--July captures the mature East Asian summer monsoon, while April--May targets the Meiyu frontal rainfall system that precedes it.}
\end{table}
\clearpage

\clearpage
\section{Supplementary Figures}\label{sec:eval_regions}

\begin{figure}[h]
\centering
\includegraphics[width=\textwidth]{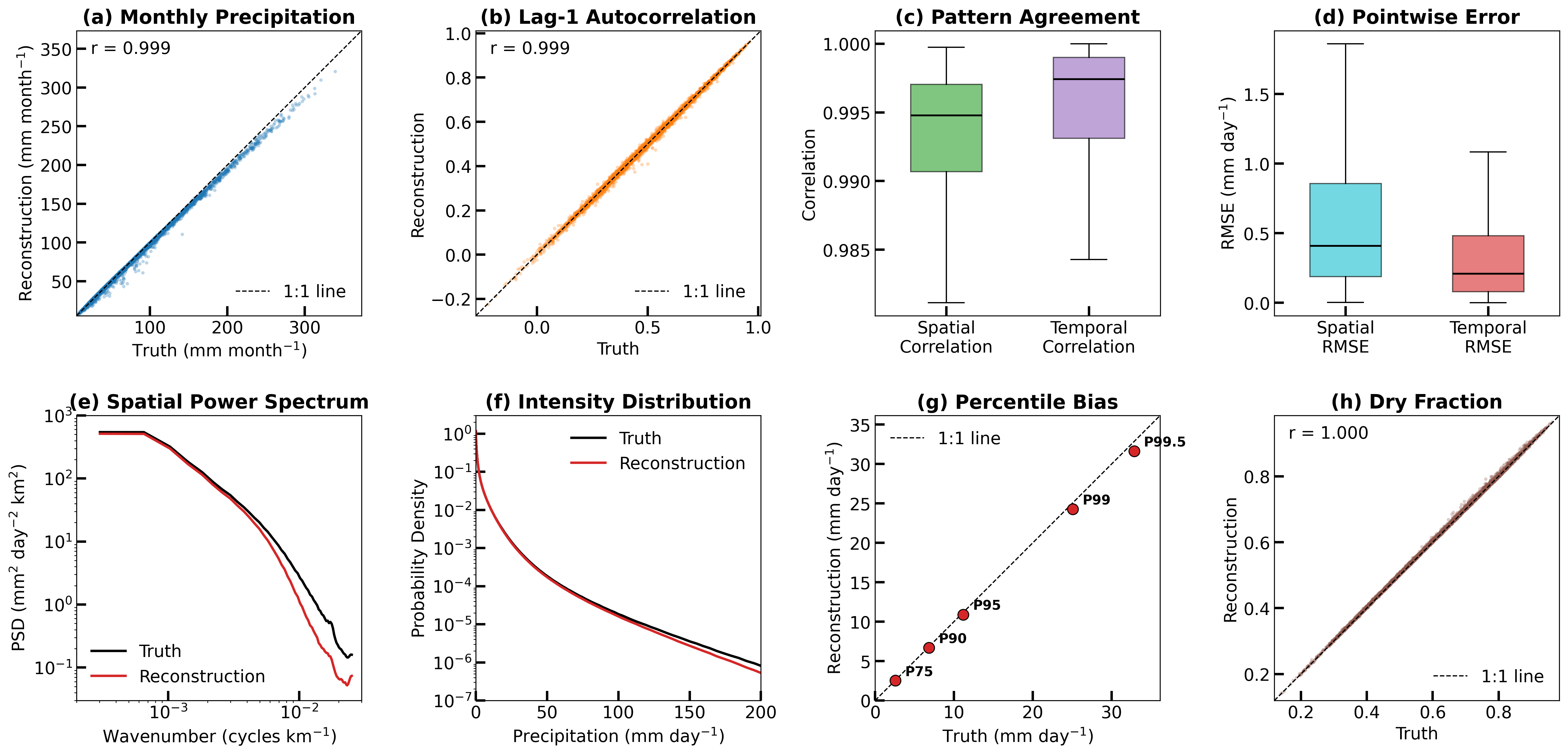}
\caption{\textbf{Diagnostic evaluation of VAE reconstruction quality.} \textbf{(a)}~Spatially averaged monthly total precipitation for reference and reconstructed fields. \textbf{(b)}~Lag-1 temporal autocorrelation from spatially averaged daily precipitation series. \textbf{(c)}~Pearson correlations computed spatially at each timestep and temporally at each grid point between reference and reconstructed fields. \textbf{(d)}~Root mean square error (RMSE), computed as in (c). \textbf{(e)}~Isotropic spatial power spectral density averaged over the test set. \textbf{(f)}~Precipitation intensity distribution across all grid points and timesteps. \textbf{(g)}~Sample-averaged precipitation percentiles from P75 to P99.5. \textbf{(h)}~Dry fraction, defined as the proportion of values below $1~\text{mm}~\text{day}^{-1}$. Dashed lines in (a), (b), (g), and (h) indicate the 1:1 reference. Boxes in (c) and (d) show the median and interquartile range; whiskers extend to $1.5\times$ the interquartile range.}
\label{fig:vae_metrics}
\end{figure}

\clearpage
\begin{figure}[tbp]
\centering
\includegraphics[width=\textwidth]{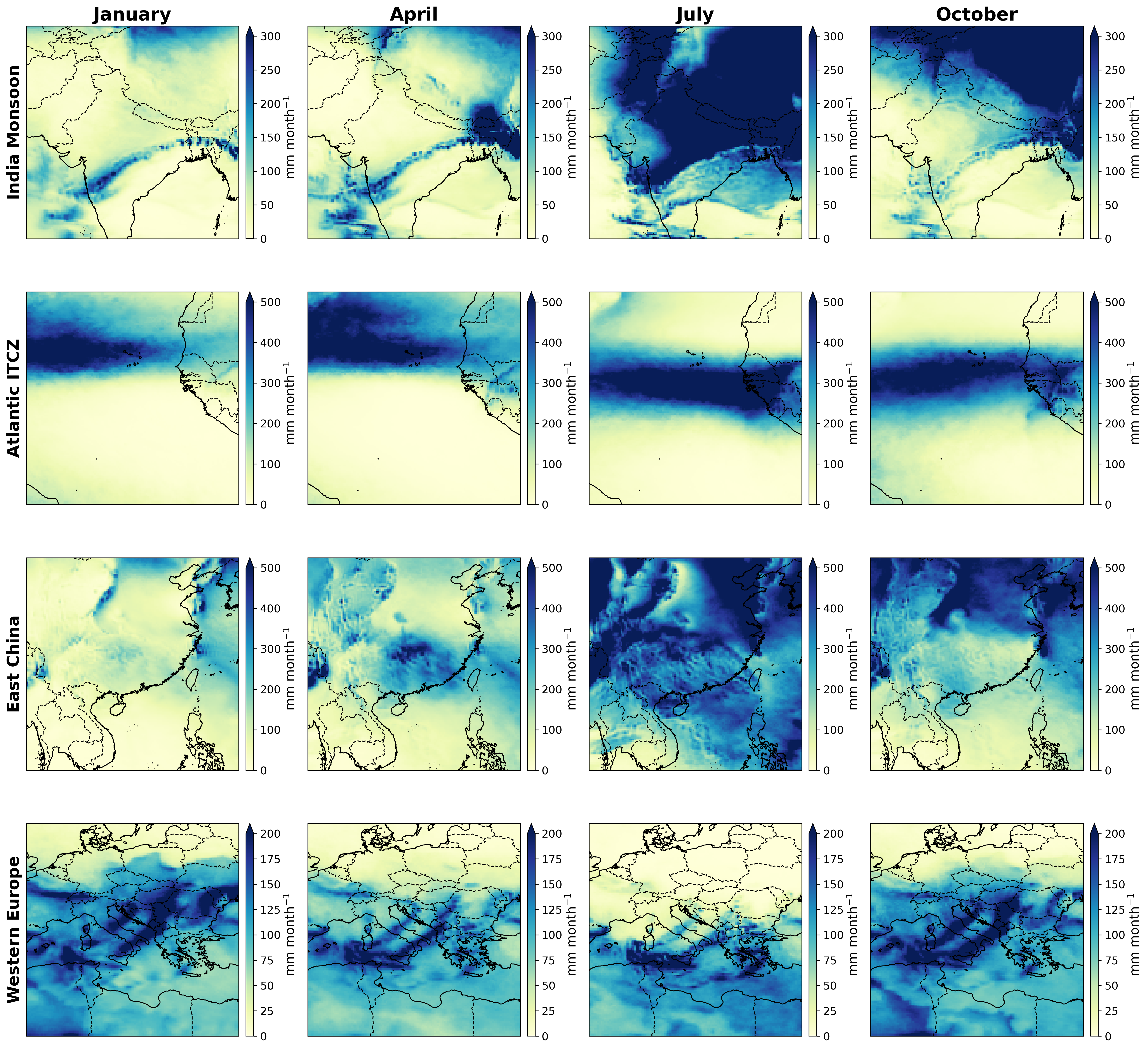}
\caption{95th-percentile sample of monthly total precipitation from 50 samples generated by the conditional diffusion prior.}
\label{fig:cond_prior_p95}
\end{figure}
\clearpage

\clearpage
\begin{figure}[tbp]
\centering
\includegraphics[width=\textwidth]{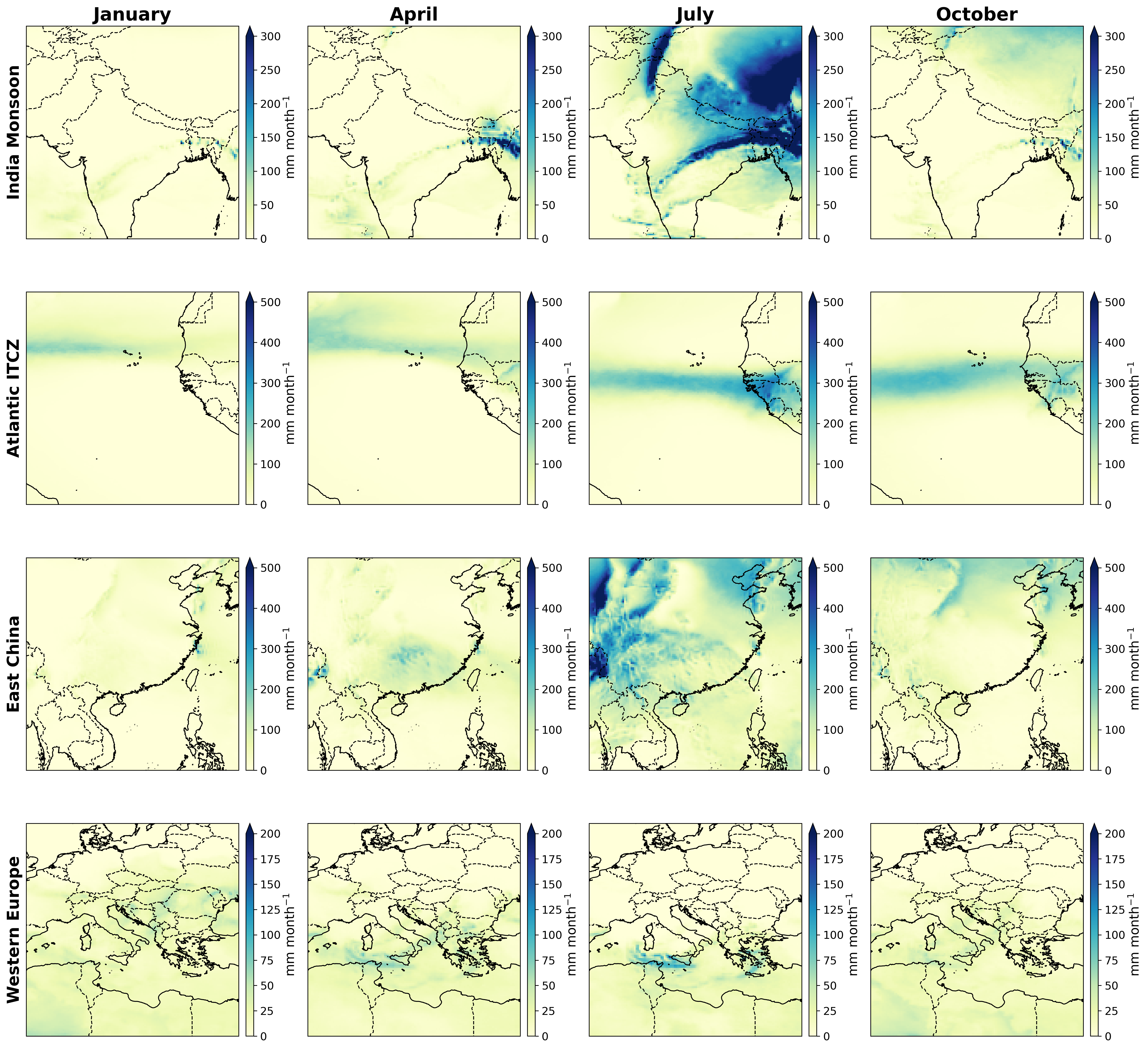}
\caption{5th-percentile sample of monthly total precipitation from the same ensemble as Fig.~\ref{fig:cond_prior_p95}.}
\label{fig:cond_prior_p05}
\end{figure}
\clearpage

\clearpage
\begin{figure}[tbp]
\centering
\includegraphics[width=\textwidth]{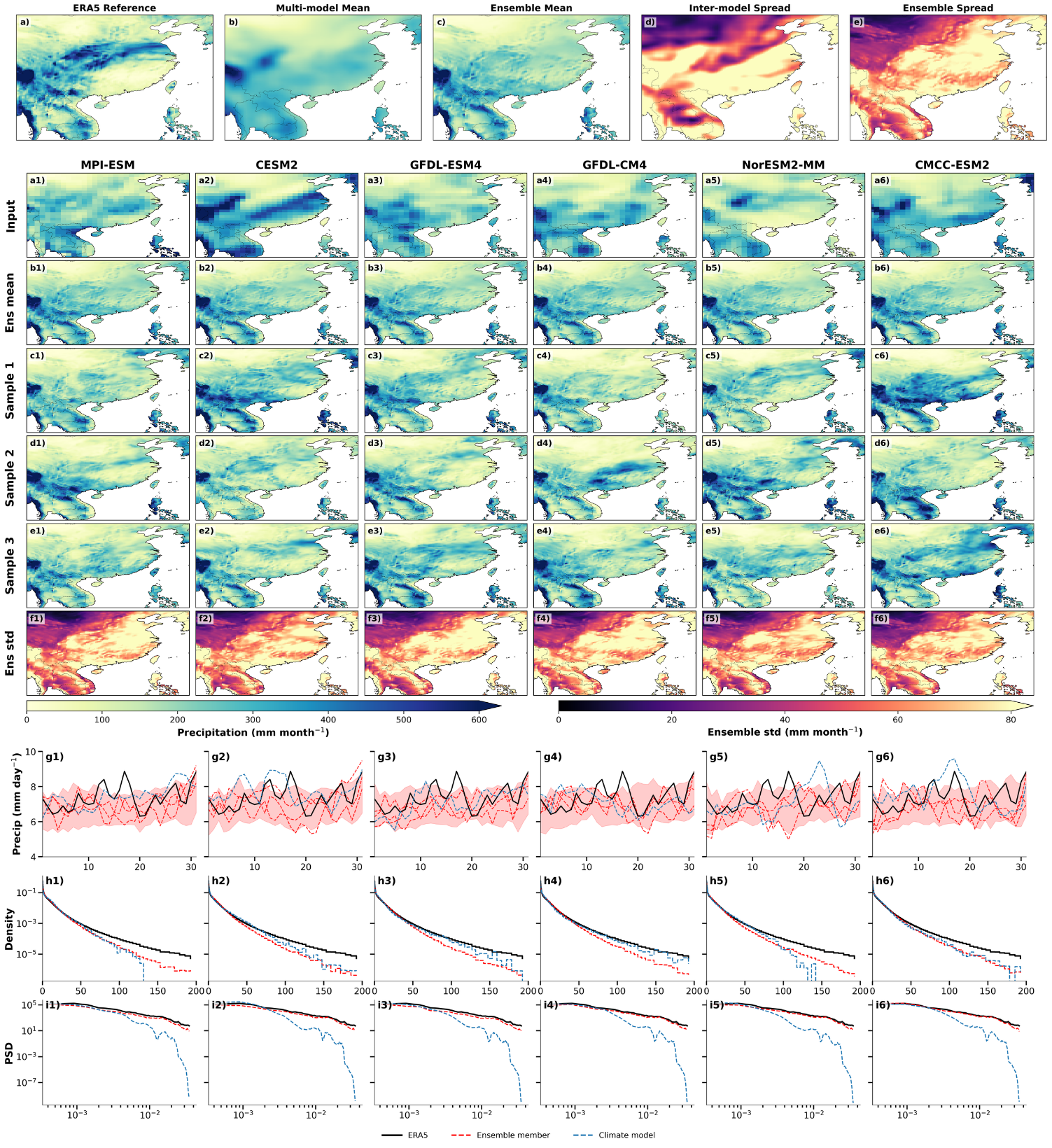}
\caption{\textbf{Stochastic downscaling of six CMIP6 climate models over East China, compared with ERA5 for July 2005--2010.} \textbf{(a--e)} Monthly precipitation totals for July 2007: \textbf{(a)} ERA5 reference at $0.25^\circ$ resolution; \textbf{(b)} mean of the six climate model inputs after bilinear upsampling to the ERA5 grid; \textbf{(c)} mean of all generated ensemble members across the six models; \textbf{(d)} pixelwise standard deviation across the six climate model inputs; \textbf{(e)} pixelwise standard deviation across all generated ensemble members. \textbf{(a1--a6)} Coarse climate model inputs at approximately $1^\circ$ resolution for July 2007, one per model. \textbf{(b1--b6)} Per-model ensemble means of the 50 Longwang downscaling samples. \textbf{(c1--c6, d1--d6, e1--e6)} Three individual ensemble members per model. \textbf{(f1--f6)} Per-model standard deviation across the 50 ensemble members. \textbf{(g1--g6)} Daily spatially-averaged precipitation climatology for each model, averaged across the six July periods. The black line shows ERA5 climatology; the blue dashed line shows the climate model's raw daily climatology; the red dashed lines show three individual ensemble member climatologies; the red shaded band shows the 5--95\% range across the 50 ensemble members. \textbf{(h1--h6)} Probability density of pixel-level precipitation values per model, pooled across all six years; for the ensemble, the curve is computed using one randomly selected member per year. \textbf{(i1--i6)} Isotropic spatial power spectral density of monthly precipitation fields per model, averaged over the six years; climate model inputs are bilinearly interpolated to the ERA5 grid.}
\label{fig:mpi_china_2}
\end{figure}
\clearpage

\clearpage
\bibliography{reference}